\documentclass[
aps,%
11pt,%
final,%
notitlepage,%
oneside,%
twocolumn,%
nobibnotes,%
nofootinbib,%
superscriptaddress,%
noshowpacs,%
centertags]%
{revtex4}

\usepackage{multirow}
\renewcommand{\baselinestretch}{1.0}
\begin{document}

\selectlanguage{english}


\title{New  H$\boldsymbol{\alpha}$ Flux Measurements in Nearby Dwarf Galaxies}

\author {\firstname{S.~S.}~\surname{Kaisin}}
\email{skai@sao.ru} 

 \author {\firstname{I.~D.}~\surname{Karachentsev}}
 \email{ikar@sao.ru}
\affiliation{Special  Astrophysical  Observatory,  Russian  Academy  of  Sciences, Russia}

\received{September 2, 2014}  \revised{September 8, 2014}
\onecolumngrid
{\scriptsize ISSN 1990-3413, Astrophysical Bulletin, 2014, Vol. 69, No. 4, pp. 390--408. c Pleiades Publishing, Ltd., 2014.
Original Russian Text c S.S. Kaisin, I.D. Karachentsev, 2014, published in Astrofizicheskii Byulleten,
 2014, Vol. 69, No. 4, pp. 413--431.}

\begin{abstract}
We present the emission H$\alpha$ line images for 40 galaxies of
the Local Volume based on the observations at the  6-meter BTA
telescope. Among them there are eight satellites of the Milky Way
and Andromeda (M\,31) as well as two companions to M\,51.  The
measured $H\alpha$ fluxes of the galaxies are used to determine
their integral (SFR) and specific (sSFR) star formation rates. The
values of $\log{\rm sSFR}$
 for the observed galaxies lie in the range of $(-9, -14)$~[yr$^{-1}]$.
 A comparison of SFR estimates derived from the $H\alpha$
flux and from the ultraviolet FUV flux yields evidence that two
blue compact galaxies MRK\,475 and LV\,J1213+2957 turn out to be
at a sharp peak of their star-burst activity.

\end{abstract}

\maketitle

\section{INTRODUCTION}

This work is a continuation of papers~[1--8] determining the rate
of star formation in the galaxies of the Local Volume measuring
their integral flux in the  emission H$\alpha$ Balmer line. Our
program provides measurements of the H$\alpha$ flux at the 6-m
telescope of the Special Astrophysical Observatory for all
galaxies of the northern sky with distances within $D\leq11$~Mpc
which were not covered by observations at other telescopes.

The latest version of the Updated Nearby Galaxy Catalog
(UNGC)~[9] contains~869 objects with distance estimates within   $D<11$~Mpc
or with the line-of-sight velocities relative to the centroid of the Local Group
$V_{\rm LG}<600$~km\,s$^{-1}$. During  2014 this sample has been
supplemented with forty new galaxies included in the database~[10], which is available at
 {\tt http://www.sao.ru/lv/lvgdb}. In total the current database of the
Local Volume contains    H$\alpha$ flux measurements  for approximately
550 galaxies, where the fluxes for more than 350 galaxies were measured with the
6-m  BTA telescope.

The   H$\alpha$ line images of  nearby galaxies  make it possible
to study the distribution of the centers of star formation that
formed in the galaxies over the past 10--30~myr. A comparison of
the H$\alpha$-emission map of the galaxy with the distribution of
atomic and molecular hydrogen in it allows a detailed study of the
conditions under which a transformation of the gas component in
the galaxy into stars takes place.

Of particular interest in this respect are the dwarf galaxies with
their shallow gravitational wells, owing to which  dwarf systems
easily lose their gas component under the effect of the supernova
burst or during the motion through the halo of hot gas around a
massive neighboring galaxy. Being sensitive to the environment,
dwarf galaxies are often used as ``test particles'' for the
analysis of the dynamical evolution of both galaxy clusters and
the  population of the general field.

Below we present the H$\alpha$ images, the H$\alpha$ flux values,
and the  star formation rate (SFR) estimates for forty galaxies of
the Local Volume and one object, KKR\,9, which turned out to be an
interstellar cirrus. All but two  galaxies  are dwarf systems of
late or early types.

\section{OBSERVATIONS AND DATA REDUCTION}

The images in the  H$\alpha$ line and in the neighboring continuum
were obtained in the period from February to December 2013. The
observations were carried out at the prime focus of the BTA with
the SCORPIO focal reducer~[11] equipped with a $2048 \times 2048$
pixel CCD, in the
 $2\times2$ binning mode.  The optical system provided a field
of view of   \mbox{$6\farcm1\times6\farcm1$} at the scale of $0\farcs185$
per pixel.

The images in the \mbox{H$\alpha$ + [N\,II]} lines   were obtained
using a narrowband interference filter with the effective
wavelength of  6555~\AA \ and the full width at half maximum of
75~\AA. The images in the continuum were taken with midband
filters  SED\,607 (\mbox{$\lambda_e=6063$~\AA},
$\Delta\lambda=167$~\AA) and SED\,707
(\mbox{$\lambda_e=7036$~\AA},\linebreak
\mbox{$\Delta\lambda=207$~\AA}). Owing to a small range of
line-of-sight velocities of the galaxies, all the objects were
observed with one and the same filter set. The typical total
exposure in the H$\alpha$ line was 1200~s.

For processing the observed data, standard procedures were used:
bias subtraction, flat-field correction, removing cosmic ray hits,
sky background subtraction. The images in the continuum were
normalized to the H$\alpha$ images using the non-overexposed stars
and  were then subtracted. Integral H$\alpha$ fluxes of galaxies
were measured from  the
  \mbox
{H$\alpha$ images} with subtracted continuum. To calibrate the
flux,   the images of the  spectrophotometric standard stars~[12]
were used which were observed at the same night with the objects.
A typical H$\alpha$ flux logarithm  measurement error in our
images is $\pm0.1$~[8].

\section{RESULTS OF OBSERVATIONS}

The mosaic image  of forty-one  observed objects is given in  the
Appendix, where the left-hand-side images correspond to the total
exposure in the H$\alpha$  line and the continuum, and the
right-hand-side images    are the
 H$\alpha$ line images after subtracting the continuum.  The image scale  and
their  \mbox{``north--east''} orientation are marked at the bottom
of the right-hand-side images. After the continuum subtraction
many images reveal residual ``caverns'' caused by the saturation
of stellar images in  high-brightness stars. Another reason may be
an abnormal color index of stars or different image quality in the
H$\alpha$ filter and in the continuum. In fact, it is the presence
of such residual components that determines the accuracy limit
when detecting the \mbox{H$\alpha$ flux}  of the  galaxy,
especially if it has a low surface brightness and large size or is
located in a dense stellar field  at a low galactic latitude.

Following~[13], we determine the integral star formation rate   in
the galaxy using the relation
$$\log{\rm SFR}=\log F_c({\rm H}\alpha) + 2\log D +8.98,$$
where $F_c({\rm H}\alpha)$ is  the integral flux in the  H$\alpha$
line in the units of erg\,cm$^{-2}\,$s$^{-1}$, corrected for the
absorption of light in the Galaxy according to~[14], \mbox{$D$ is}
the distance in Mpc, and SFR is expressed in $M_{\odot}$/year. We
have ignored the internal absorption of light in the galaxy as
well as the contribution of the emission  [N\,II] doublet, since
both effects are negligibly small for dwarf galaxies~[15, 16].

The table contains the main data on the observed galaxies. Its
columns show: (1)~the name of the galaxy, (2)~equatorial
coordinates for the epoch (J2000.0), (3)~integral apparent
$B$~magnitude from the  UNGC catalog~[9], (4)~morphological type
by the
 de~Vaucouleurs scale, (5)~distance to the galaxy in Mpc according to the UNGC
catalog data~[9], (6)~logarithm of the observed
galaxy flux in the  H$\alpha$ line in the units of
 erg\,cm$^{-2}\,$s$^{-1}$,
(7)~logarithm of the integral star formation rate in
$M_{\odot}$/year, (8)~specific star formation rate  ${\rm sSFR} =
{\rm SFR}/M^*$, where $M^*$ is the stellar mass of the galaxy from
catalog~[9],   determined from the  $K$-band luminosity.
The last column shows for comparison the integral
starburst rate of the galaxy~[16]
$$\log{\rm SFR}=\log F_c({\rm FUV}) + 2\log D - 6.78$$
determined by its flux in the far ultraviolet\linebreak
(\mbox{$\lambda_e=1539$~\AA}, ${\rm FWHM}=269$~\AA),  measured by
the
 GALEX satellite~[17], corrected for the extinction of light in the Galaxy.

\renewcommand{\baselinestretch}{0.8}
\begin{table*}
\setcaptionmargin{0mm} \onelinecaptionstrue
\captionstyle{nonumber} \caption{\centerline{Parameters of the observed galaxies}}
\medskip
\begin{tabular}{l|c|c|r|c|r|r|r|rl}
\hline
\multicolumn{1}{c|}{Name} &   RA,     Dec   &  $B_t$  &  $T$ &  $D$   & \multicolumn{1}{c|}{$\log F$} & $\log{\rm SFR}$& $\log{\rm sSFR}$  &\multicolumn{2}{c}{$\log{\rm SFR}_{\rm FUV}$}\\
\hline

And\,XXXII~$=$C~as\,III &  003559.4+513335  & 13.7 & $-3$ & 0.78  & $<-15.28$ & $<-6.35$ & $<-13.92$ & & \multicolumn{1}{c}{--}\\
P\,And\,AS-48           &  005928.2+312910  & 20.5 & $-3$ & 0.82  & $<-15.17$ & $<-6.30$ & $<-10.99$ & $<$&$-6.37$\\
DGSAT-I                 &  011736.2+333134  & 18.2 & $-2$ & 8.0   & $<-15.23$ & $<-4.39$ & $<-11.95$ & &$-3.86$\\
Segue\,2                &  021916.0+201031  & 16.2 & $-3$ & 0.03  & $<-15.37$ & $<-9.14$ & $<-13.02$ & &$-7.92$\\
And\,XXXIII~$=$~Pers\,I &  030123.6+405918  & 15.5 & $-2$ & 0.79  & $<-15.28$ & $<-6.38$ & $<-13.19$ & &  \multicolumn{1}{c}{--}\\
NGC\,1400               &  033930.8--184117 & 11.9 & $-3$ & 24.9  &  $-12.49$ &  $-0.66$ &  $-11.65$ & &$-1.21$\\
NGC\,1592               &  042940.8--272431 & 14.5 & 10   & 9.1   &  $-12.21$ &  $-1.28$ &  $-9.46$  & &$-0.93$\\
ESO\,489-056            &  062617.0--261556 & 15.7 & 10   & 4.99  &  $-13.19$ &  $-2.75$ &  $-10.24$ & &  \multicolumn{1}{c}{--}  \\
HIPASS\,J0705--20       &  070545.0--205930 &   -- & 10   & 7.2   &  $-14.46$ &  $-3.18$ &    --     & &  \multicolumn{1}{c}{--}\\
LV\,J0737+4724          &  073728.5+472433  & 18.1 & 10   & 15.7  &  $-14.30$ &  $-2.83$ &  $-10.44$ & &$-2.48$\\
LV\,J0812+4836          &  081239.5+483645  & 17.1 &  9   & 12.4  &  $-13.92$ &  $-2.70$ &  $-10.40$ & &$-2.46$\\
KUG\,0821+3201          &  082505.0+320103  & 16.8 & 10   & 8.2   &  $-13.61$ &  $-2.77$ &  $-10.21$ & & \multicolumn{1}{c}{--} \\
AGC\,182595             &  085112.1+275248  & 17.2 &  9   & 9.04  &  $-13.78$ &  $-2.85$ &  $-10.22$ & &$-2.87$\\
LV\,J0913+1937          &  091339.0+193708  & 17.4 & 10   & 4.4   &  $-13.82$ &  $-3.51$ &  $-10.18$ & &$-3.36$  \\
AGC\,198508             &  092257.0+245648  & 17.8 & 10   & 10.4  &  $-13.52$ &  $-2.47$ &  $-9.71$  & &  \multicolumn{1}{c}{--}\\
MCG\,+09-16-010         &  092317.0+515822  & 16.2 &  8   & 7.4   &  $-13.43$ &  $-2.70$ &  $-10.34$ & &$-2.42$\\
UGC\,5047               &  092849.6+513338  & 16.0 &  7   & 19.7  &  $-13.82$ &  $-2.22$ &  $-10.94$ & &  \multicolumn{1}{c}{--}\\
KUG\,0937+4800          &  094019.6+474638  & 17.0 &  9   & 7.9   &  $-13.83$ &  $-3.04$ &  $-10.32$ & &$-2.71$\\
LV\,J1000+5022          &  100025.5+502245  & 17.1 &  9   & 8.0   &  $-13.68$ &  $-2.89$ &  $-10.13$ & & \multicolumn{1}{c}{--} \\
LV\,J1000+3032          &  100036.5+303210  & 18.1 & 10   & 7.1   &  $-14.58$ &  $-3.88$ &  $-10.65$ & &$-3.39$\\
LV\,1021+0054            &  102138.9+005400  & 17.5 & 10   & 6.8   &  $-13.71$ &  $-3.02$ &  $-10.03$ & &$-2.78$\\
Leo\,P                  &  102145.1+180517  & 17.2 & 10   & 2.0   &  $-13.69$ &  $-4.08$ &  $-10.11$ & & \multicolumn{1}{c}{--}\\
LV\,J1030+0607          &  103044.3+060738  & 16.8 & 10   & 7.8   &  $-14.01$ &  $-3.22$ &  $-10.59$ & &$-2.73$\\
AGC\,731457             &  103155.8+280133  & 16.8 & 10   & 11.12 &  $-13.73$ &  $-2.63$ &  $-10.31$ & &$-2.27$\\
UMa\,I                  &  103452.8+515512  & 15.3 & $-3$ & 0.10  & $<-15.28$ & $<-8.28$ & $<-13.16$ & $<$&$-8.36$\\
KUG\,1033+366B          &  103617.6+362531  & 17.0 &  9   & 8.1   &  $-13.60$ &  $-2.78$ &  $-10.09$ & &$-2.74$\\
LV\,J1052+3628          &  105205.5+362836  & 17.9 & 10   & 9.2   &  $-14.40$ &  $-3.48$ &  $-10.53$ & &$-2.93$\\
LV\,J1213+2957          &  121348.4+295732  & 17.4 & 10   & 2.7   &  $-13.38$ &  $-3.53$ &  $-9.72$  & &$-3.92$\\
KKH\,78                 &  121744.5+332043  & 17.7 & 10   & 4.7   & $<-15.30$ & $<-4.96$ & $<-11.51$ & &$-4.95$\\
DDO\,120                &  122115.0+454841  & 13.6 &  9   & 7.1   &  $-15.15$ &  $-4.46$ &  $-13.17$ & &$-1.96$  \\
GR\,34                  &  122207.6+154757  & 16.0 & 10   & 9.29  &  $-14.39$ &  $-3.45$ &  $-11.31$ & &$-2.82$\\
KDG\,215                &  125540.5+191233  & 16.9 & 10   & 4.83  &  $-15.17$ &  $-4.80$ &  $-11.70$ & &$-3.33$\\
CVn\,I                  &  132803.5+333321  & 13.9 & $-3$ & 0.22  & $<-15.32$ & $<-7.62$ & $<-13.73$ & $<$&$-7.67$\\
UGCA\,361               &  133236.2+494949  & 16.7 & $-1$ & 8.0   & $<-15.26$ & $<-4.47$ & $<-12.57$ & &$-4.12$\\
MCG\,+08-25-028         &  133644.8+443557  & 15.9 & 10   & 7.7   &  $-13.30$ &  $-2.52$ &  $-10.23$ & &$-2.54$\\
KKR\,9                  &  142705.0+224125  &   -- &  --  &  --   & $<-15.24$ & $<-5.51$ &    --     & & \multicolumn{1}{c}{--}     \\
MRK\,475                &  143905.4+364822  & 16.4 &  9   & 9.2   &  $-12.41$ &  $-1.50$ &  $-9.14$  & &$-2.24$  \\
UGC\,9660               &  150109.3+444153  & 14.4 &  9   & 7.4   &  $-12.62$ &  $-1.88$ &  $-10.25$ & &$-1.81$  \\
UGC\,9992               &  154147.8+671515  & 15.4 & 10   & 7.3   &  $-13.04$ &  $-2.30$ &  $-10.19$ & &$-2.02$  \\
Segue\,3                &  212131.0+190702  & 15.9 & $-3$ & 0.02  & $<-15.37$ & $<-9.86$ & $<-13.06$ & $<$&$-9.67$\\
And\,XXXI~$=$~Lac\,I     &  225816.3+411728  & 14.0 & $-3$ & 0.76  & $<-15.28$ & $<-6.42$ & $<-13.77$ & &  \multicolumn{1}{c}{--}\\

\hline
 \end{tabular}
 \end{table*}
\renewcommand{\baselinestretch}{1.00}

\section{FEATURES OF SOME  OBSERVED GALAXIES}

With the exception of two late-type  spiral galaxies  MCG\,+09-16-010 (Sm) and UGC\,5047 (Sdm), all the remaining observed
objects are dwarf galaxies of late (Ir, Im, BCD) or early (dSph) type. Note the most remarkable
among them.

{\it And\,XXXII = Cas\,III, And\,XXXIII = Perseus\,I,
And\,XXXI = Lacerta\,I.} These three dSph companions of
M\,31 were recently detected within the\linebreak
\mbox{$3\pi$~Pan-STARRS1}-survey~[18, 19]. All   three objects do
not reveal any signs of emission. The table    lists for them only
the upper limits of the integral   H$\alpha$~flux. We should,
however, note that the angular diameters of these nearby    dwarf
systems are quite large:  $13\farcm0$~(Cas\,III),
$3\farcm4$~(Pers\,I), and $8\farcm4$~(Lac\,I),  and in two cases
they exceed the field of view of our detector.

{\it P\,And\,AS-48.} A distant globular cluster at the far
periphery of M\,31~[20].

{\it DGSAT-I.}  A dwarf spheroidal galaxy in the neighborhood of
M\,31~[21]. Its radial velocity  is not known, and the distance,
measured as 8.0~Mpc, is very unreliable.

{\it Segue\,2, UMa\,I, CVn\,I, Segue\,3.} Four dShp companions of
the Milky Way, detected in~[22--25] respectively. The table lists
for them only the upper value of the H$\alpha$ flux. It should be
borne in mind that the angular dimensions of UMa\,I~($18\farcm0$)
and CVn\,I ($14\farcm3$) significantly exceed the field of view of
the SCORPIO focal reducer.

{\it NGC\,1400.}  An elliptical galaxy in the Eridanus group at
the distance of 24.9 Mpc~[26]. It formally makes it into the
sample of the Local Volume by its  line-of-sight  velocity $V_{\rm
LG}=496$~km\,s$^{-1}$.  Its peculiar line-of-sight velocity
relative to the  Eridanus group is $-1320$~km\,s$^{-1}$. The
presence of H$\alpha$~emission in its central part  is not typical
for the E galaxies in the groups, but the  GALEX data~[17]
indicate active star formation in the center of NGC\,1400.

{\it NGC\,1592.} A knotty dIr~galaxy with a high specific star
formation rate.

{\it HIPASS\,J0705--20.} This radio source with the radial
velocity $V_{\rm LG}=528$~km\,s$^{-1}$~[27]  has no optical
identification, although the magnitude of the Galactic extinction
in its direction is not too large,   \mbox{$A_B=2\fm71$}~[14]. The
H$\alpha$ image we have obtained reveals one diffuse spot with the
coordinates\linebreak \mbox{070546.95--205932.6}  and two point
knots with\linebreak the coordinates 070543.31--205935.1
and\linebreak \mbox {070545.87--205837.7},
 which can
belong to an irregular galaxy of low surface brightness.
    Radial velocity measurement for these objects will be a crucial
test for the  accuracy of   radio source identification.

{\it Leo\,P.} A very nearby  dIr galaxy with radial velocity of
$V_{\rm LG}=135$~km\,s$^{-1}$~[28]. The main  \mbox {H$\alpha$
flux}  comes from a compact H\,II~region near the center of the
object.

{\it LV\,J1213+2957.} A compact blue    H\,II region detected in
the SDSS~[29]. Having a small radial velocity, $V_{\rm
LG}=196$~km\,s$^{-1}$, it is probably part of the group of five
galaxies around NGC\,4150 at a distance of approximately 16~Mpc,
which has a collective peculiar velocity of about
$-700$~km\,s$^{-1}$~[30].

{\it KKH\,78.} An irregular galaxy of low surface brightness with no signs of H$\alpha$ emission, a probable companion of the NGC\,4395 galaxy.

{\it DDO\,120~$=$~UGC\,7408.} A bluish Magellanic-type irregular
galaxy with no H\,II regions. A very weak diffuse H$\alpha$
emission gives the 300 times lower star formation rate than that
from  the FUV flux measured by the GALEX.

{\it GR\,34~$=$~VCC\,530, KDG\,215.} Two dIr galaxies in front of
the Virgo Cluster, the distances to which were measured in~[31].
KDG\,215 shows only a very weak diffuse emission.

{\it UGCA\,361, MCG\,+08-25-028.} These are the dSph and dIr
companions of M\,51.

{\it KKR\,9.} This object proved to be not a dwarf galaxy but a
fragment of an interstellar cirrus.

{\it MRK\,475.} A blue compact galaxy of high surface brightness.
One of the most  active objects of the Local Volume.   Judging by
the specific star formation rate \mbox{$\log ({\rm sSFR}) =
-9.14$~[yr$^{-1}]$}, this dwarf system is at the peak of its
starburst activity.

{\it UGC\,9660, UGC\,9992.} Both dwarf galaxies have vigorous
centers of star formation. Their distances, determined   by the
\mbox{Tully--Fisher} relation~[32],  7.4~Mpc and 7.3~Mpc, almost
coincide with the distance  to M\,101 (7.38~Mpc). Both galaxies
can be associated with the group M\,101, although their radial
velocities $+745$~km\,s$^{-1}$ and $+638$~km\,s$^{-1}$ are
significantly larger than that of M\,101 ($+378$~km\,s$^{-1}$).

 \section{DISCUSSION AND CONCLUSIONS}

As we can see from the last column of the table,   thirty out of
forty galaxies we have observed  have   independent   SFR
estimates from the ultraviolet flux measured by the GALEX
satellite. A fairly clear correlation is noticeable between the
SFR~estimates by the H$\alpha$~flux and FUV~flux, which is smeared
in weak fluxes. For twenty galaxies with the \mbox{$\log {\rm
SFR}({\rm H}\alpha)>-4.0$} estimates,  the average difference
between the values is
$$\langle\log {\rm SFR}({\rm H}\alpha)-\log {\rm SFR}({\rm FUV})\rangle=-0.16\pm0.08$$
 with a standard deviation of 0.35.
This difference is somewhat larger than the  typical measurement error
of the H$\alpha$ flux ($\pm0.10$) logarithm   and the   FUV flux logarithm
($\pm0.04$).

The reasons of the differences between the SFR estimates based on
the   H$\alpha$ line emission   and the ultraviolet flux of the
galaxies were discussed by many authors, in particular, in~[16]
and~[32]. As it was noted in~[33], the star formation rate by the
H$\alpha$ flux  relative to the FUV flux for the dwarf galaxies of
very low luminosity may be underestimated by   ten or more times.
The empirical mutual normalization of  ${\rm SFR}({\rm
H}\alpha)\simeq {\rm SFR}({\rm FUV})$ was made by the spiral
galaxies. It is not well suited to the dwarf systems, where the
initial stellar mass function at its bright end may be
substantially different than that of the spiral galaxies. It
should also be remembered that the   SFR  estimation based on  the
H$\alpha$ flux corresponds  to the characteristic time of about
$10^7$~yr, whereas the SFR determined by the FUV flux   covers the
range of approximately $10^8$~yr. In the presence of star
formation bursts, especially noticeable in dwarf galaxies, the
${\rm SFR}({\rm H}\alpha)/{\rm SFR}({\rm FUV})$ ratio   indicates
the phase of  activity in which the galaxy dwells. For example,
blue compact galaxies MRK\,475 and LV\,J1213+2957 are obviously
observed at
 the peak of their starburst activity.  On the other hand,
dwarf galaxies DDO\,120 and KDG\,215 with bluish but aged stellar
population are at the stage of long-term relaxation after the
activity peaks. Therefore, the SFR estimate ratio  from the
H$\alpha$ and FUV fluxes can be a useful
 indicator of the flare activity phase of a dwarf galaxy.

As noted in~[34, 35], specific star formation rate $\log{\rm
sSFR}=\log({\rm SFR}/M^*)$  of the galaxies in the present epoch
has an upper limit expressed as $-9.4$~[yr$^{-1}]$.  Not less than
\mbox {98--99\%} of all galaxies are subject to this restriction.
It obviously means\pagebreak\ that a too vigorous star formation
leads to the depletion of gas reserves
 in the galaxy. Among the objects of the table there is only one
galaxy,  MRK\,475 with  $\log {\rm sSFR}=-9.14$~[yr$^{-1}]$, which
exceeds the specified limit. It should be noted, however, that the
error in the determining the stellar mass in blue compact galaxies
is quite large---about 50\%, and this Markarian galaxy may
actually not be violating the general pattern.

\begin{acknowledgments}
This work was supported by the RFBR grants
\mbox{13-02-92960-Ind-f} and 13-02-00780. The observations at the
6-meter BTA telescope were carried out with the financial support
of the Ministry of Education and Science of the Russian Federation
(state contracts No.~14.518.11.7070, 16.518.11.7073).
 \end{acknowledgments}

\onecolumngrid

\clearpage

\section*{APPENDIX}

\vbox{
\centerline{\includegraphics[angle=0,width=0.8\textwidth,clip]{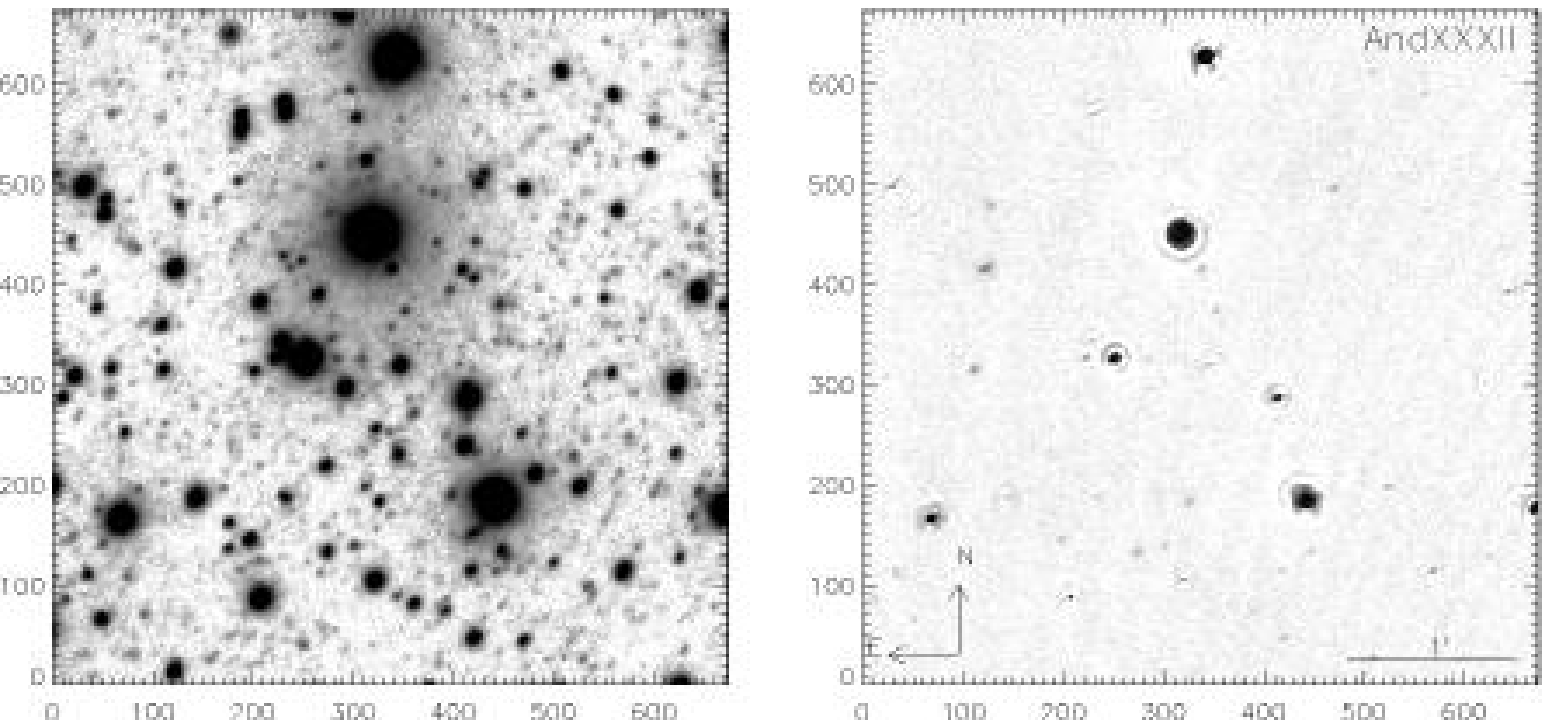}\vspace{10mm}
}}

\vbox{
 \centerline{\includegraphics[angle=0,width=0.8\textwidth,clip]{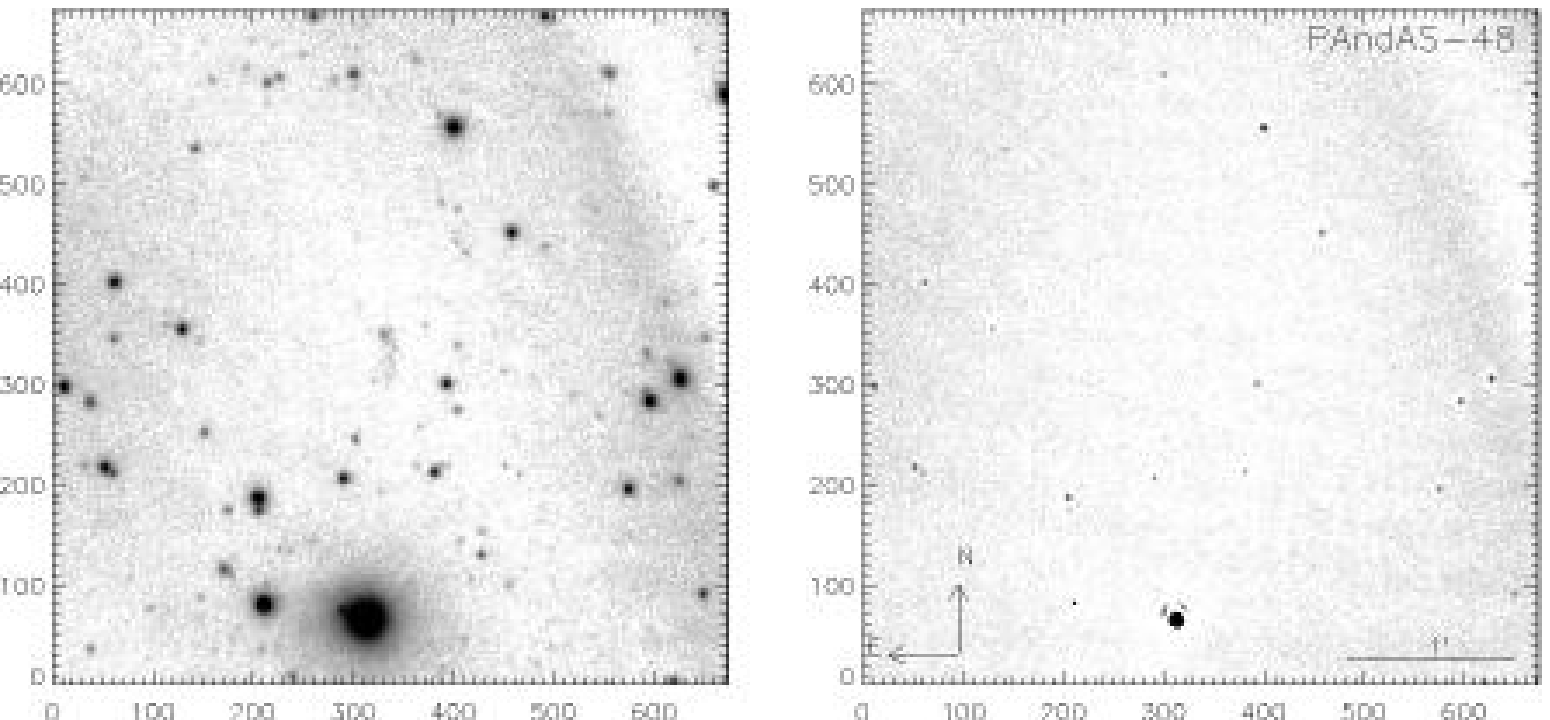}
 \vspace{2cm}
}}

{\vbox {\vspace{6mm}
\centerline{\includegraphics[angle=0,width=0.8\textwidth,clip]{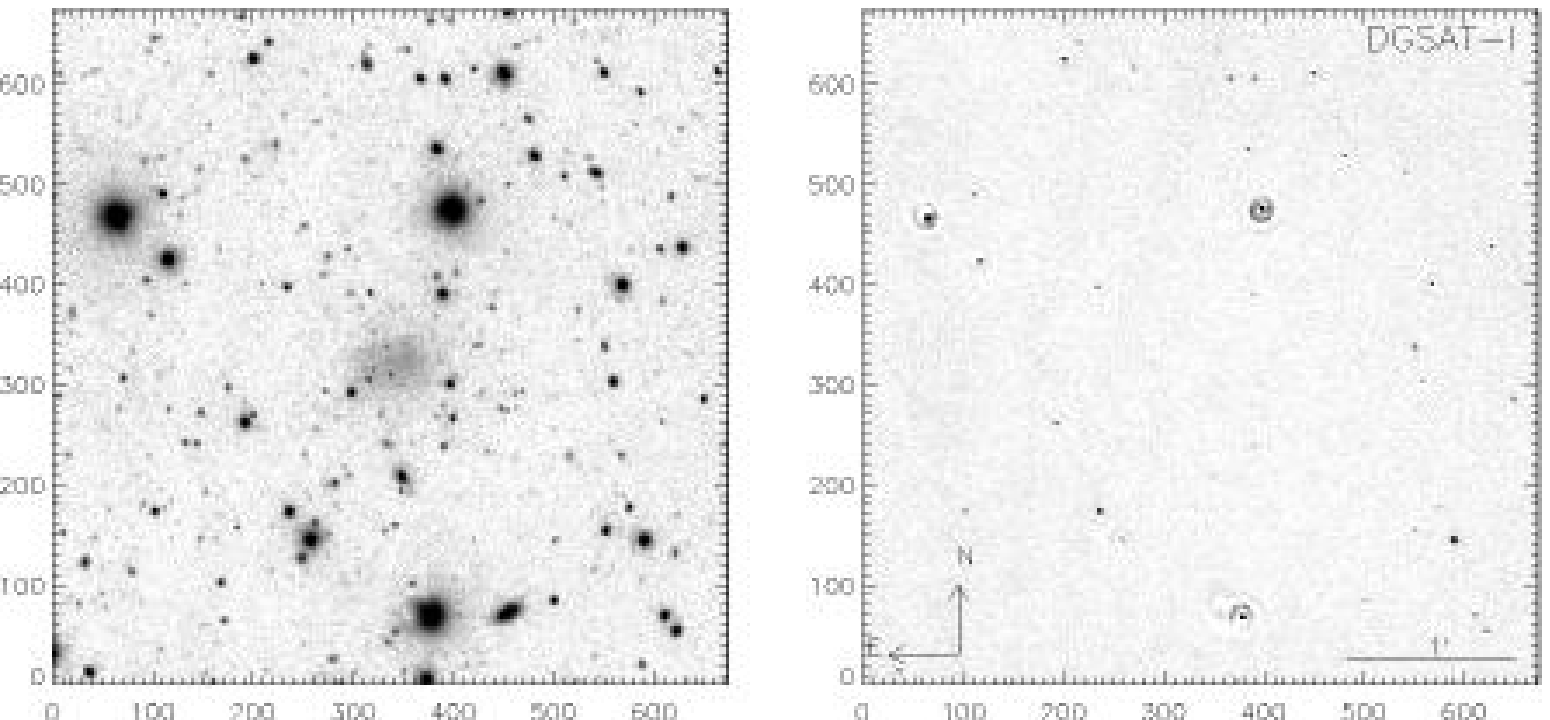}}}}
\centerline{\includegraphics[angle=0,width=0.8\textwidth,clip]{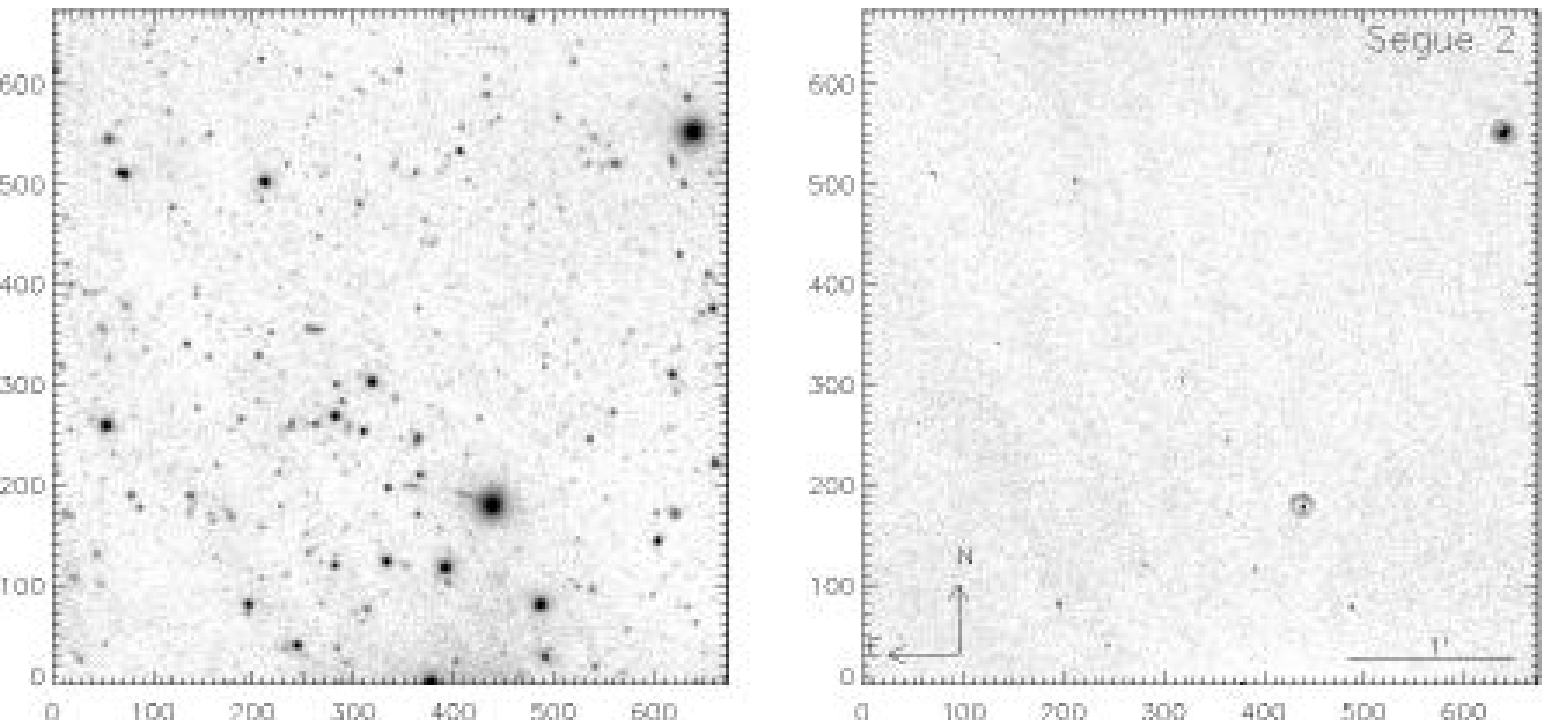}}
\centerline{\includegraphics[angle=0,width=0.8\textwidth,clip]{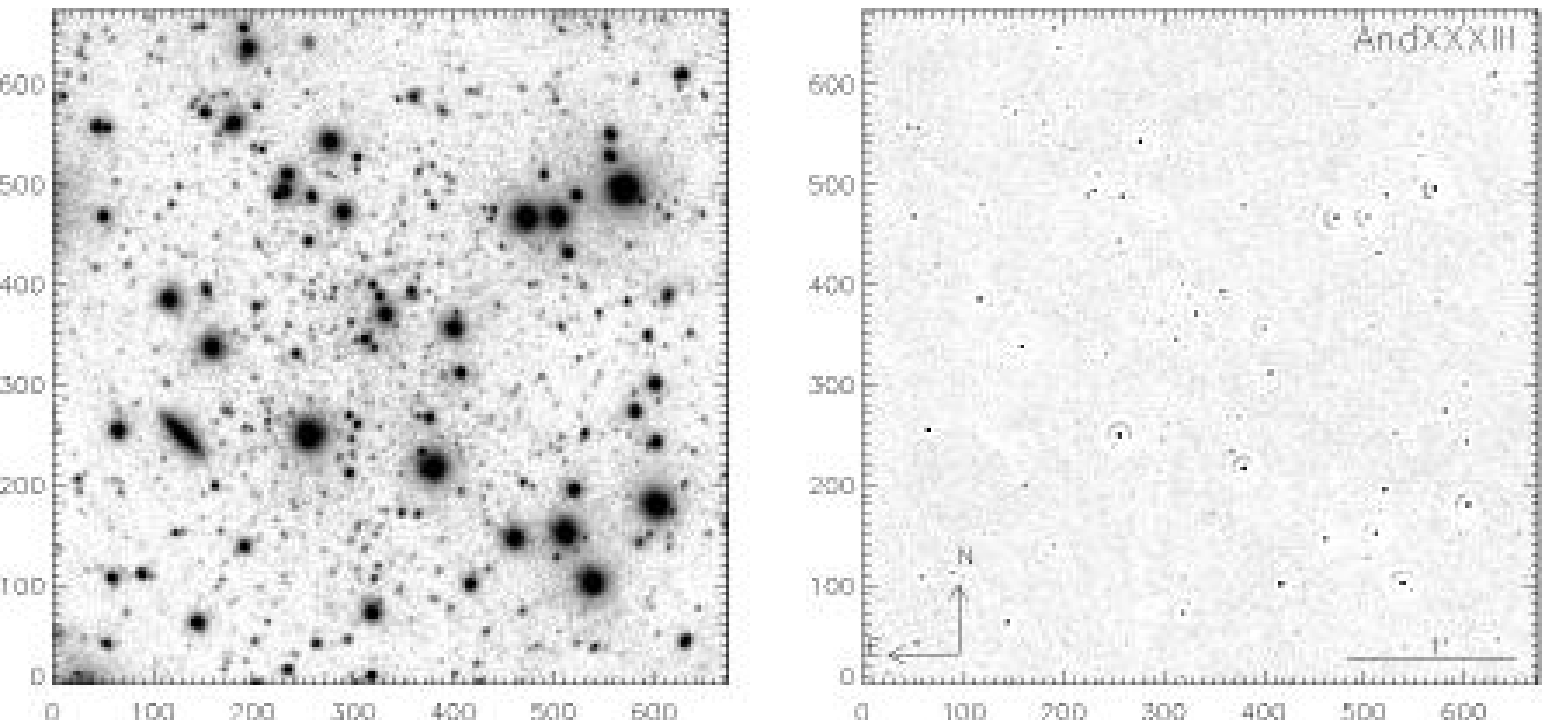}}

{\vbox {\vspace{6mm}
\centerline{\includegraphics[angle=0,width=0.8\textwidth,clip]{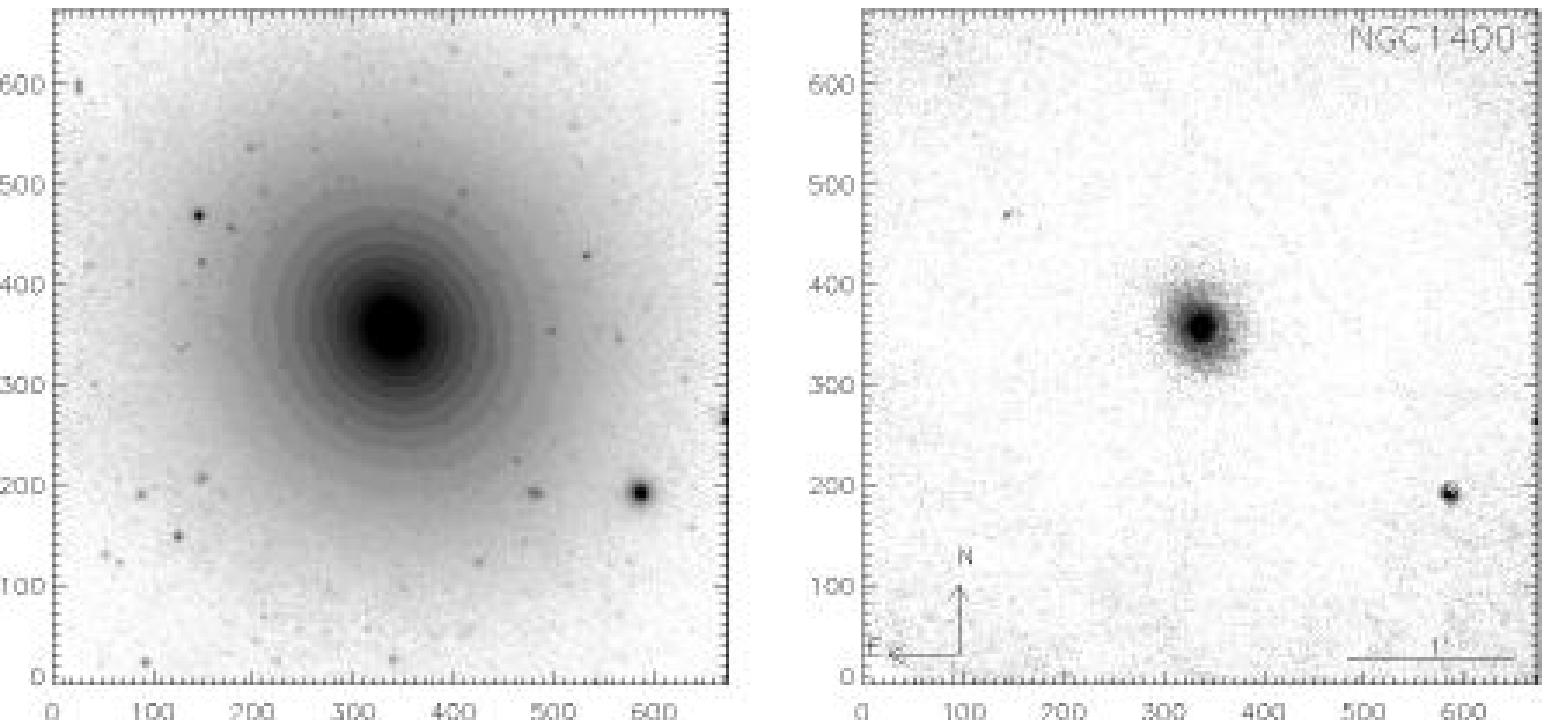}}}}
\centerline{\includegraphics[angle=0,width=0.8\textwidth,clip]{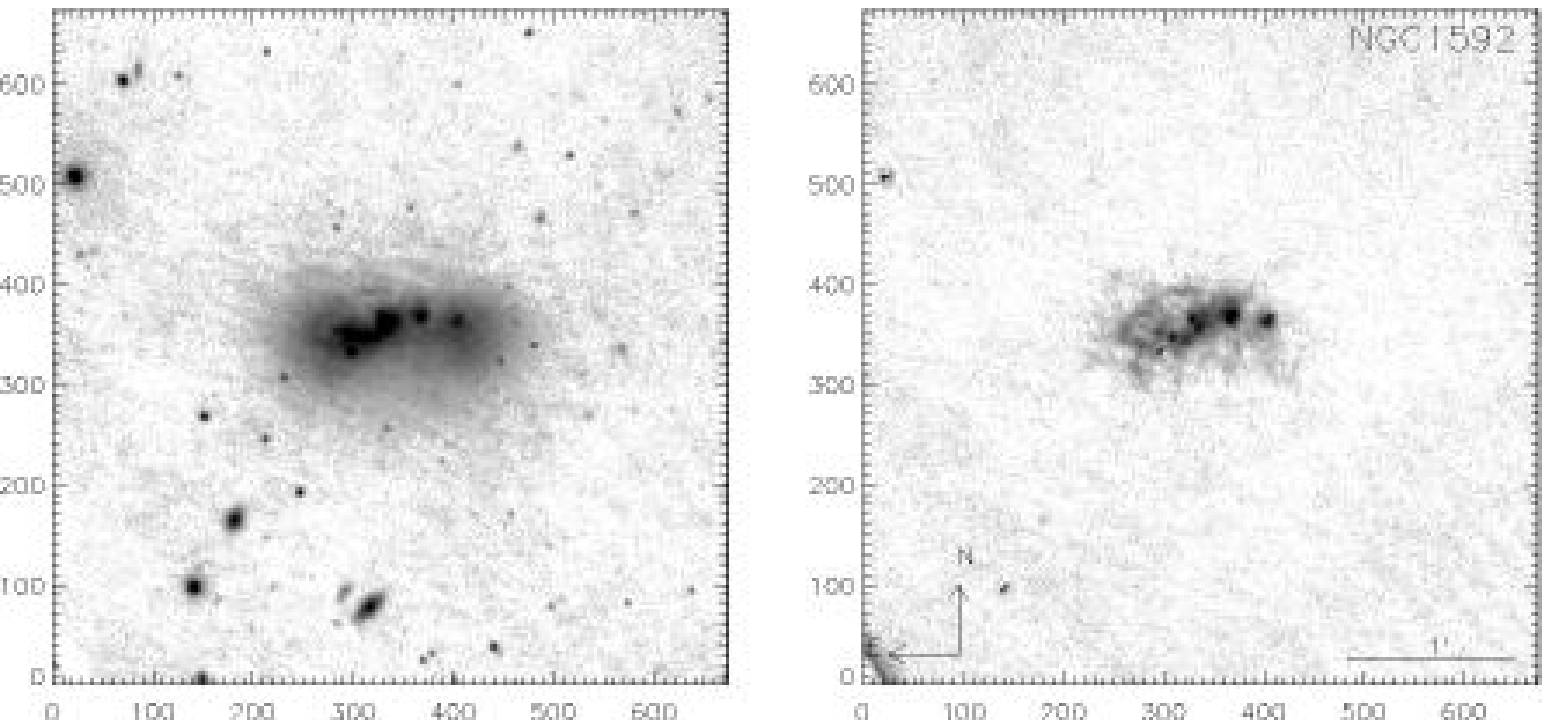}}
\centerline{\includegraphics[angle=0,width=0.8\textwidth,clip]{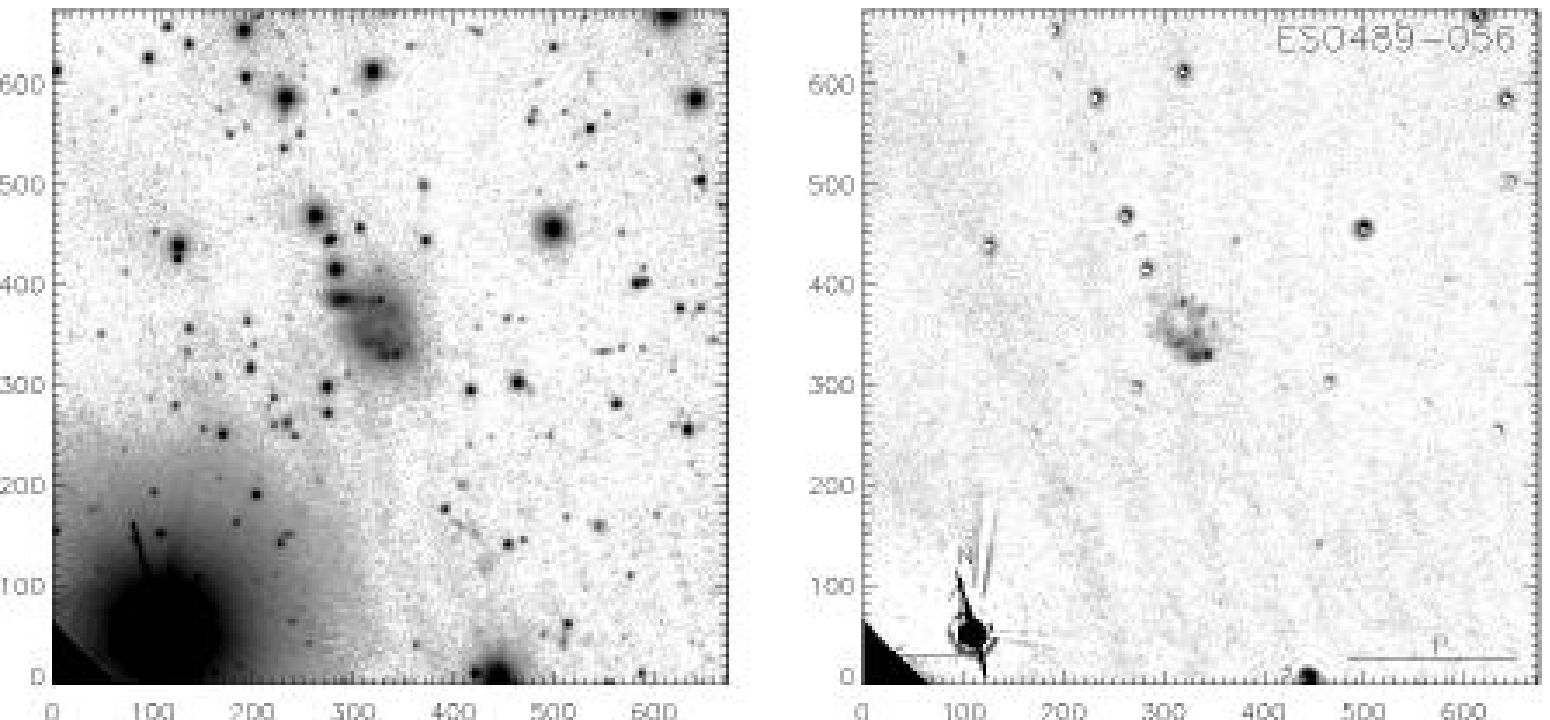}}

{\vbox {\vspace{6mm}
\centerline{\includegraphics[angle=0,width=0.8\textwidth,clip]{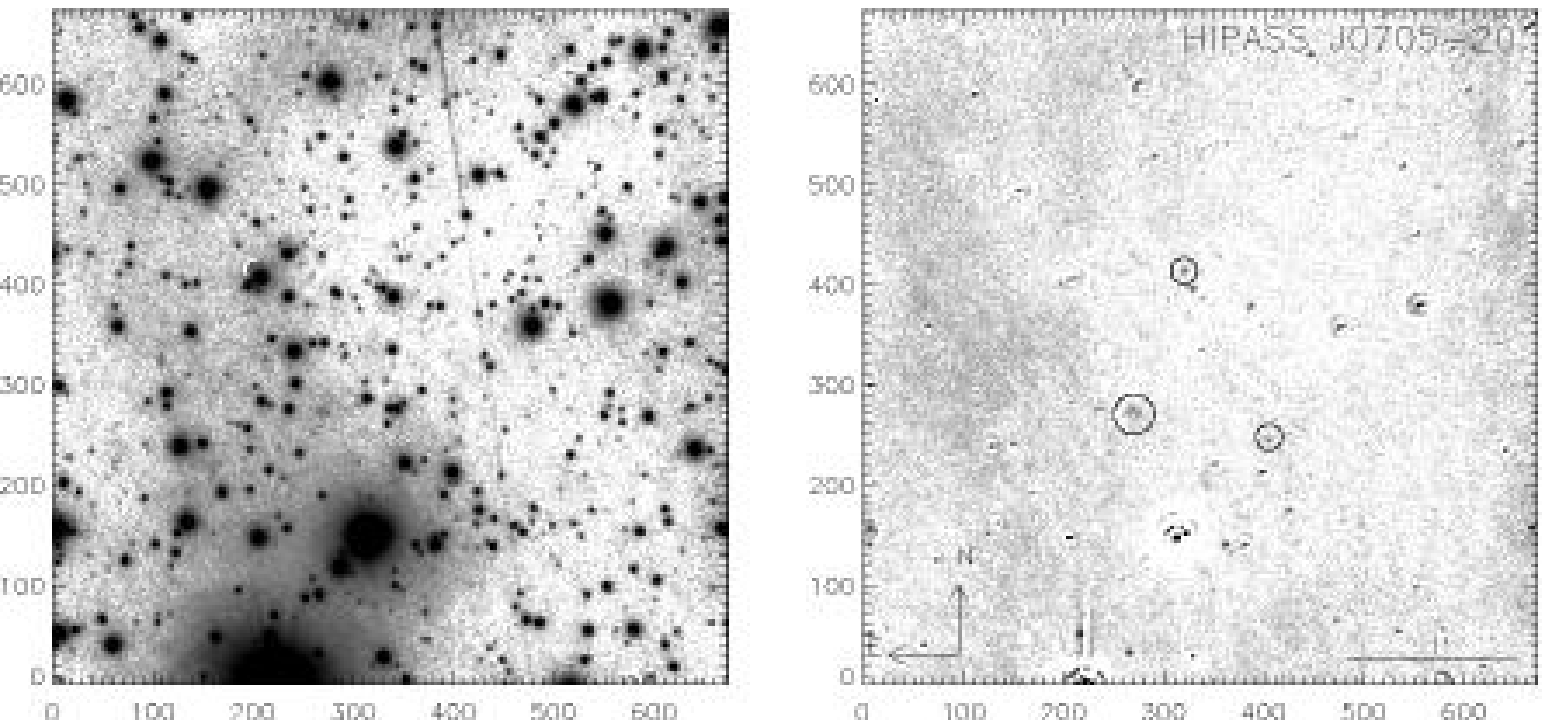}}}}
\centerline{\includegraphics[angle=0,width=0.8\textwidth,clip]{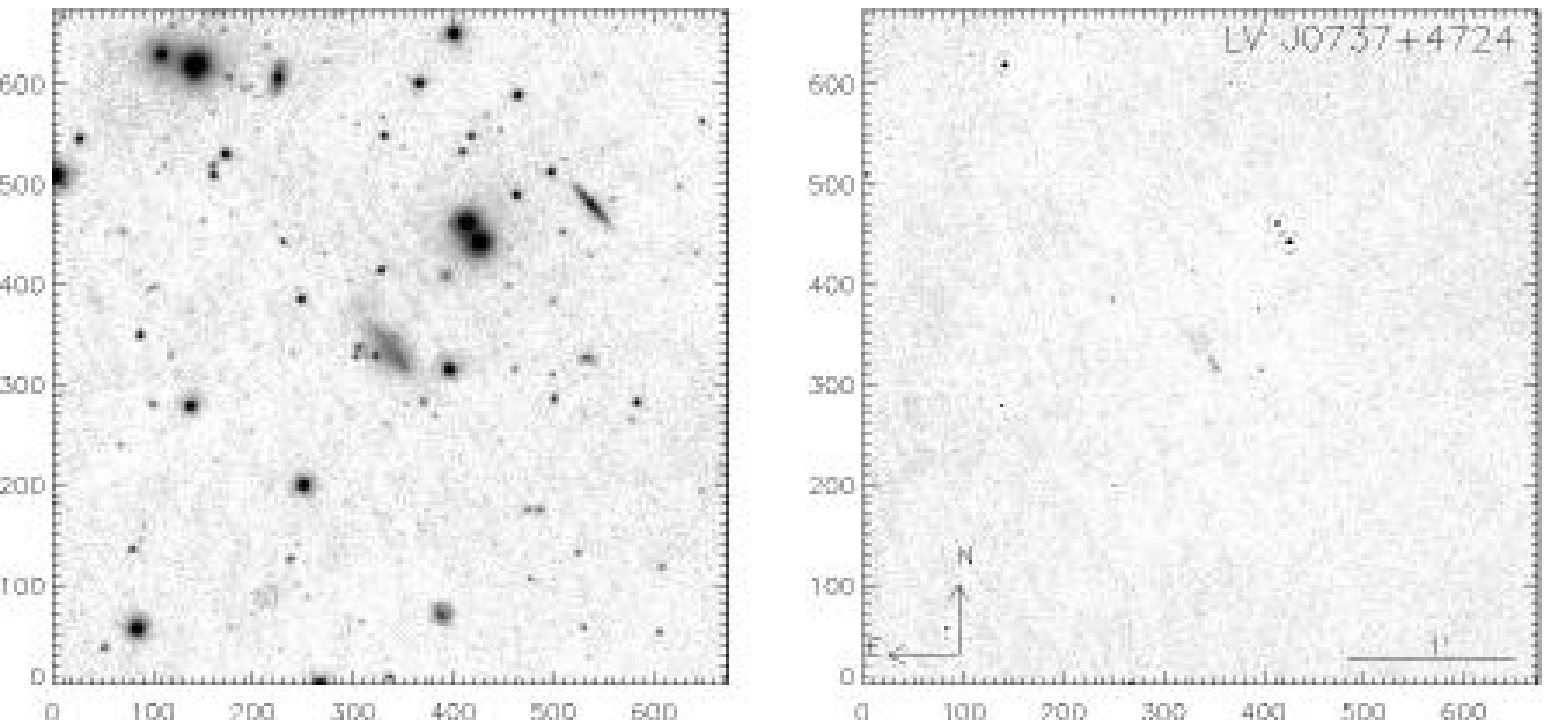}}
\centerline{\includegraphics[angle=0,width=0.8\textwidth,clip]{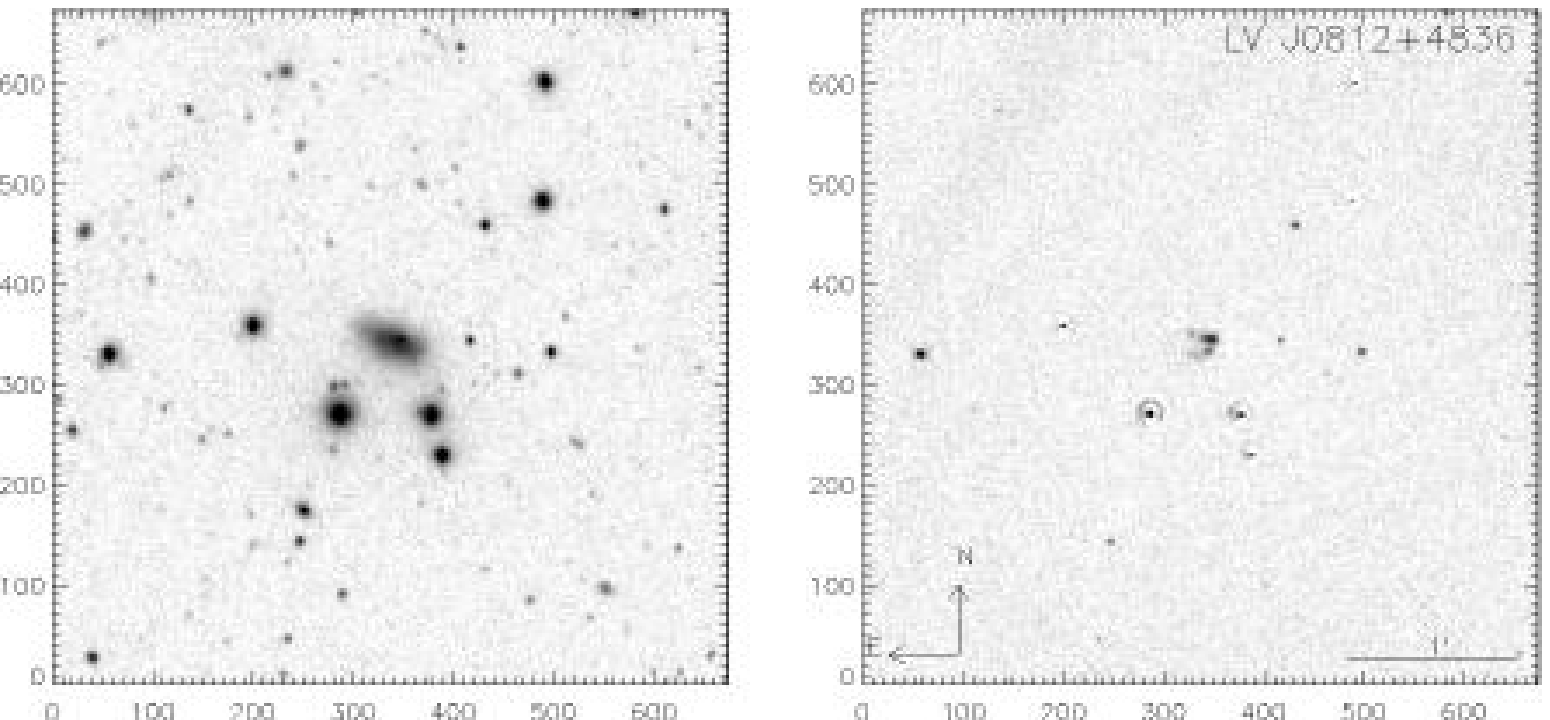}}

{\vbox {\vspace{6mm}
\centerline{\includegraphics[angle=0,width=0.8\textwidth,clip]{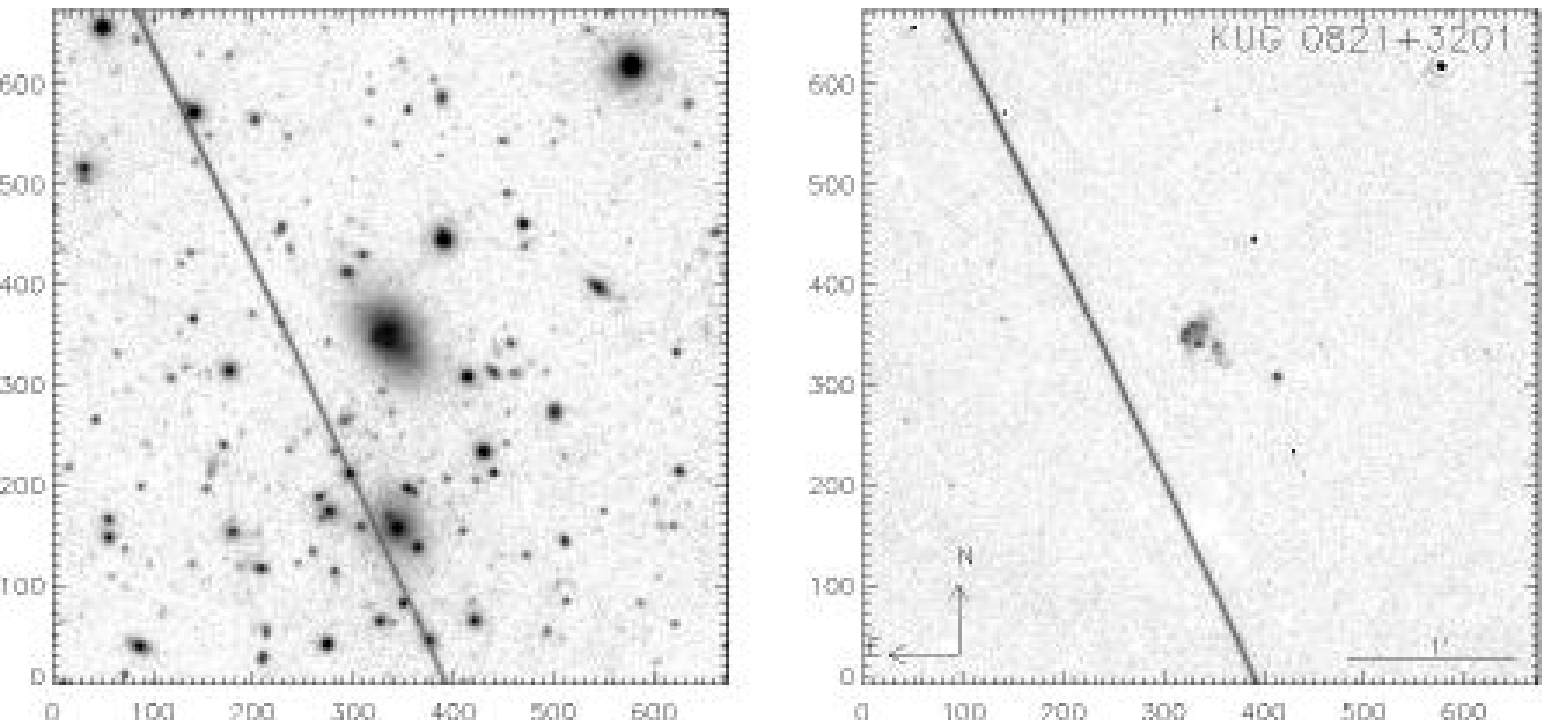}}}}
\centerline{\includegraphics[angle=0,width=0.8\textwidth,clip]{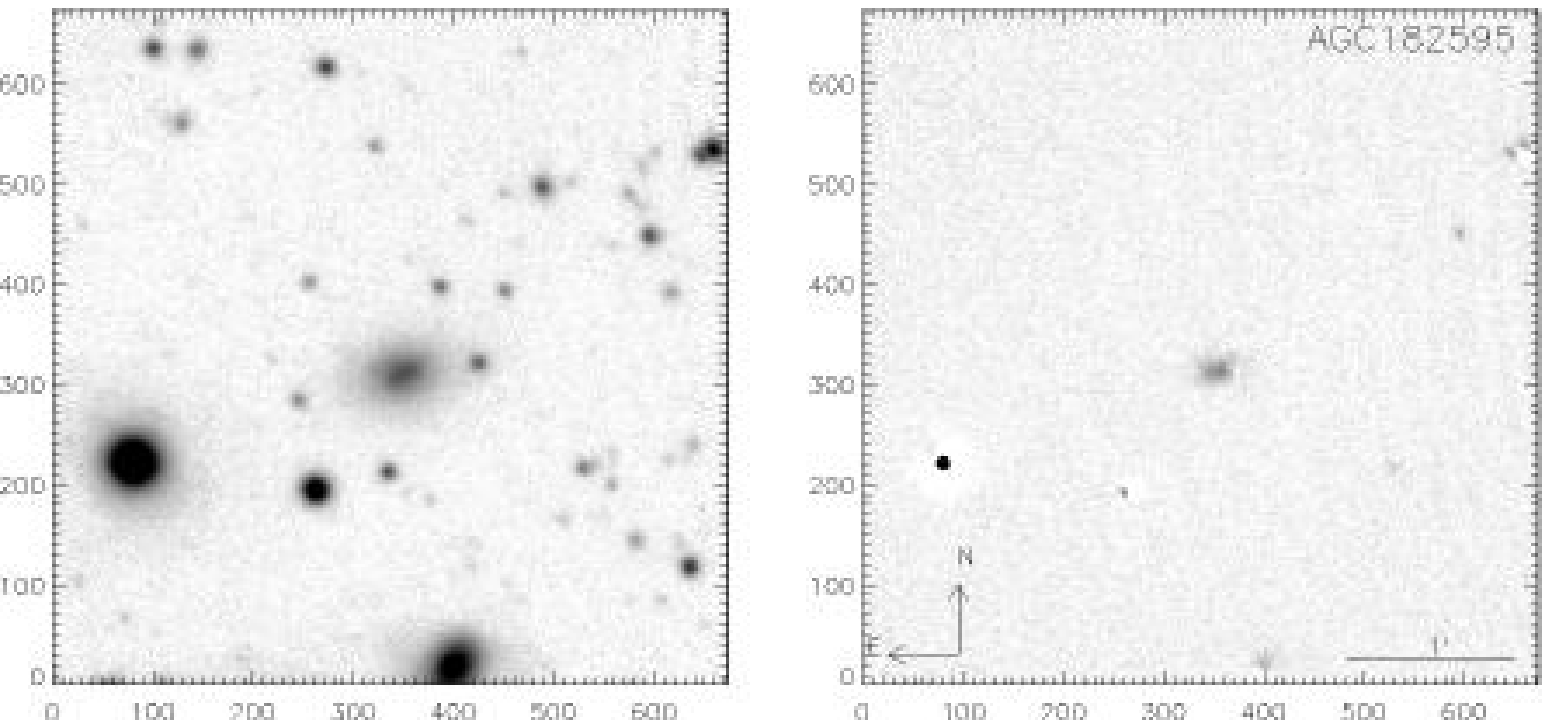}}
\centerline{\includegraphics[angle=0,width=0.8\textwidth,clip]{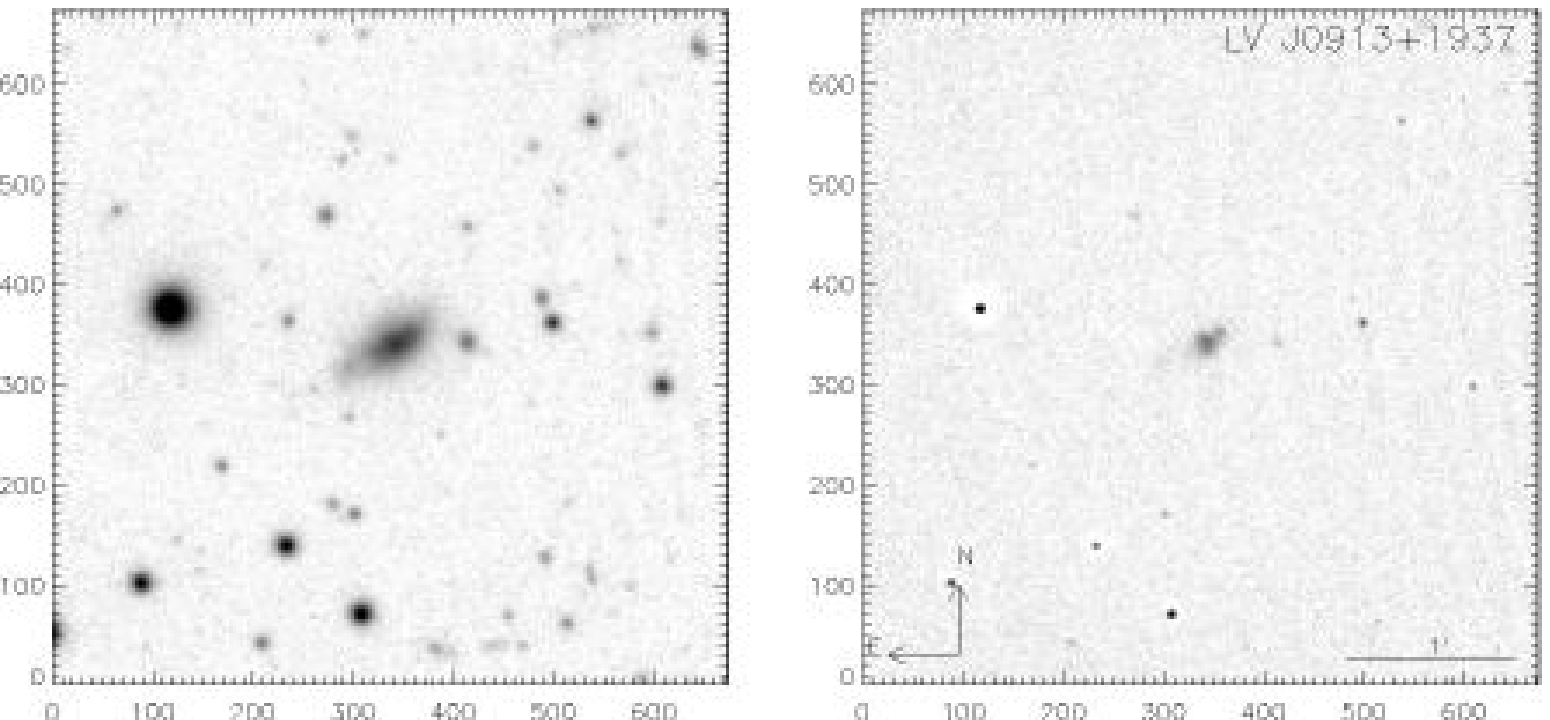}}

{\vbox {\vspace{6mm}
\centerline{\includegraphics[angle=0,width=0.8\textwidth,clip]{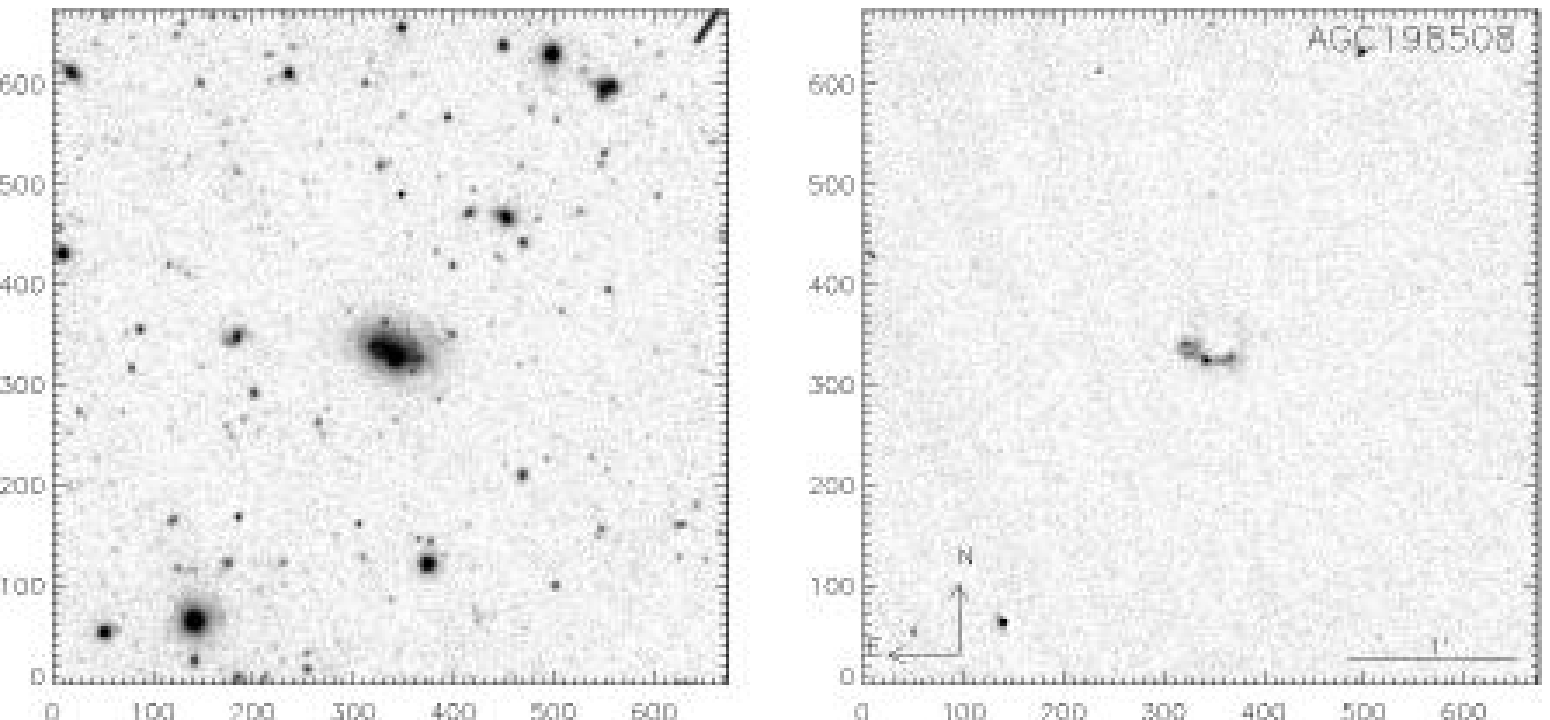}}}}
\centerline{\includegraphics[angle=0,width=0.8\textwidth,clip]{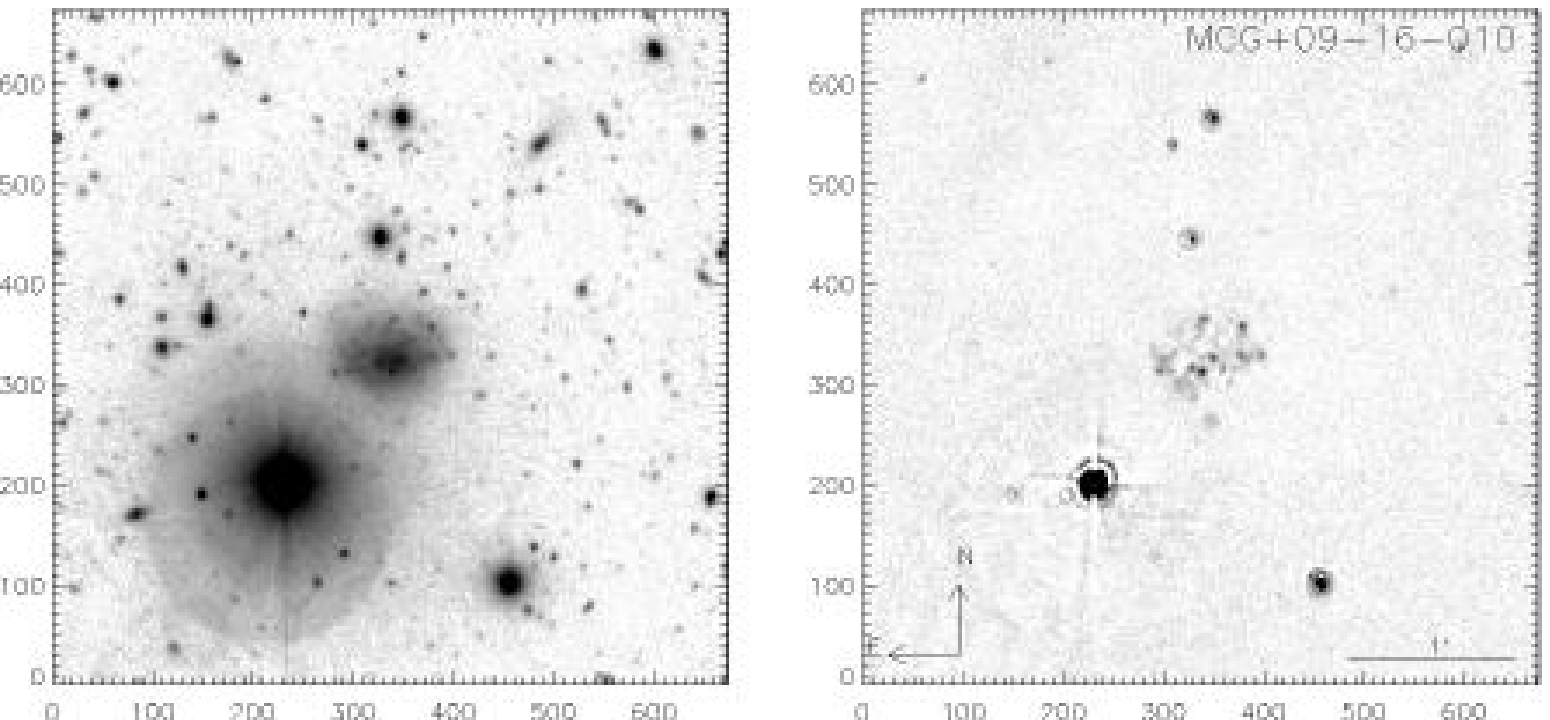}}
\centerline{\includegraphics[angle=0,width=0.8\textwidth,clip]{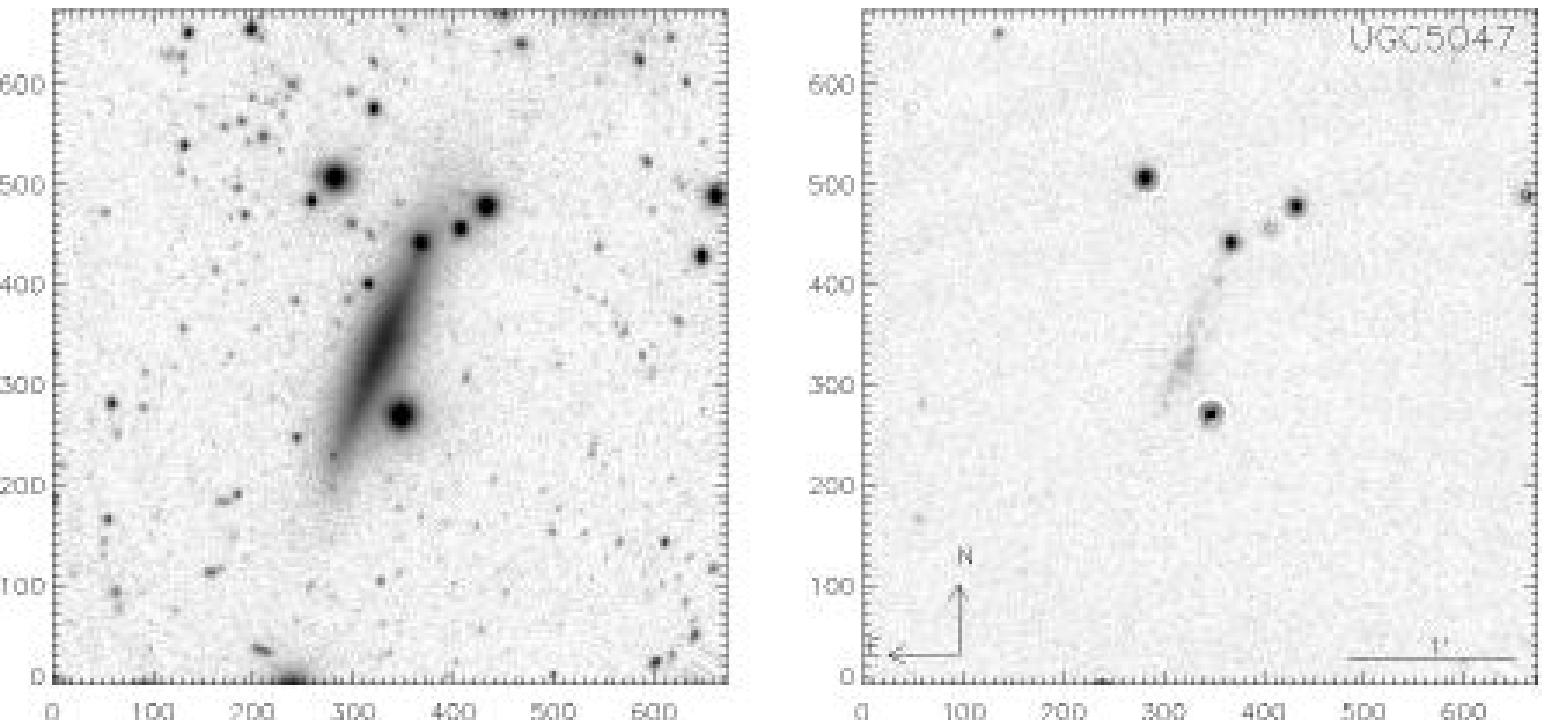}}

{\vbox {\vspace{6mm}
\centerline{\includegraphics[angle=0,width=0.8\textwidth,clip]{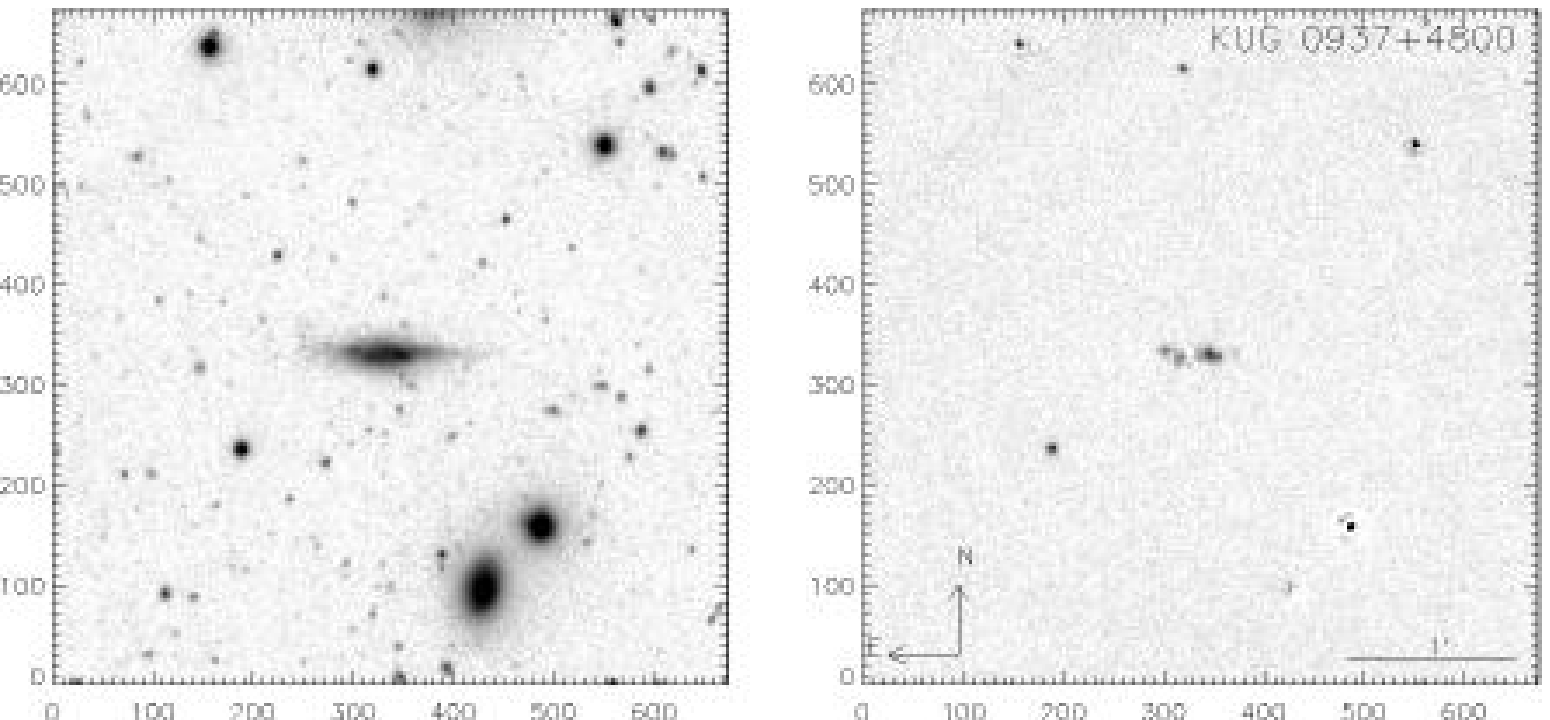}}}}
\centerline{\includegraphics[angle=0,width=0.8\textwidth,clip]{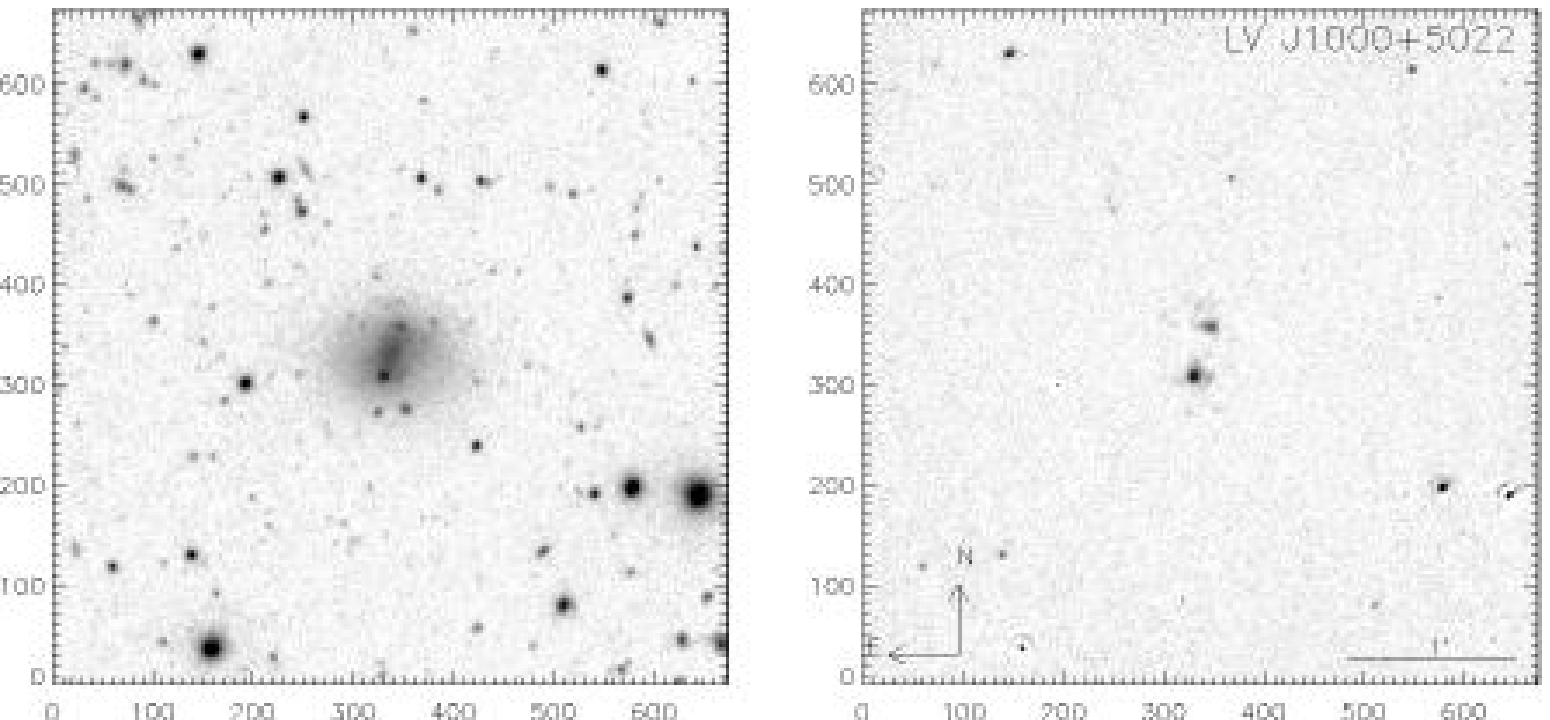}}
\centerline{\includegraphics[angle=0,width=0.8\textwidth,clip]{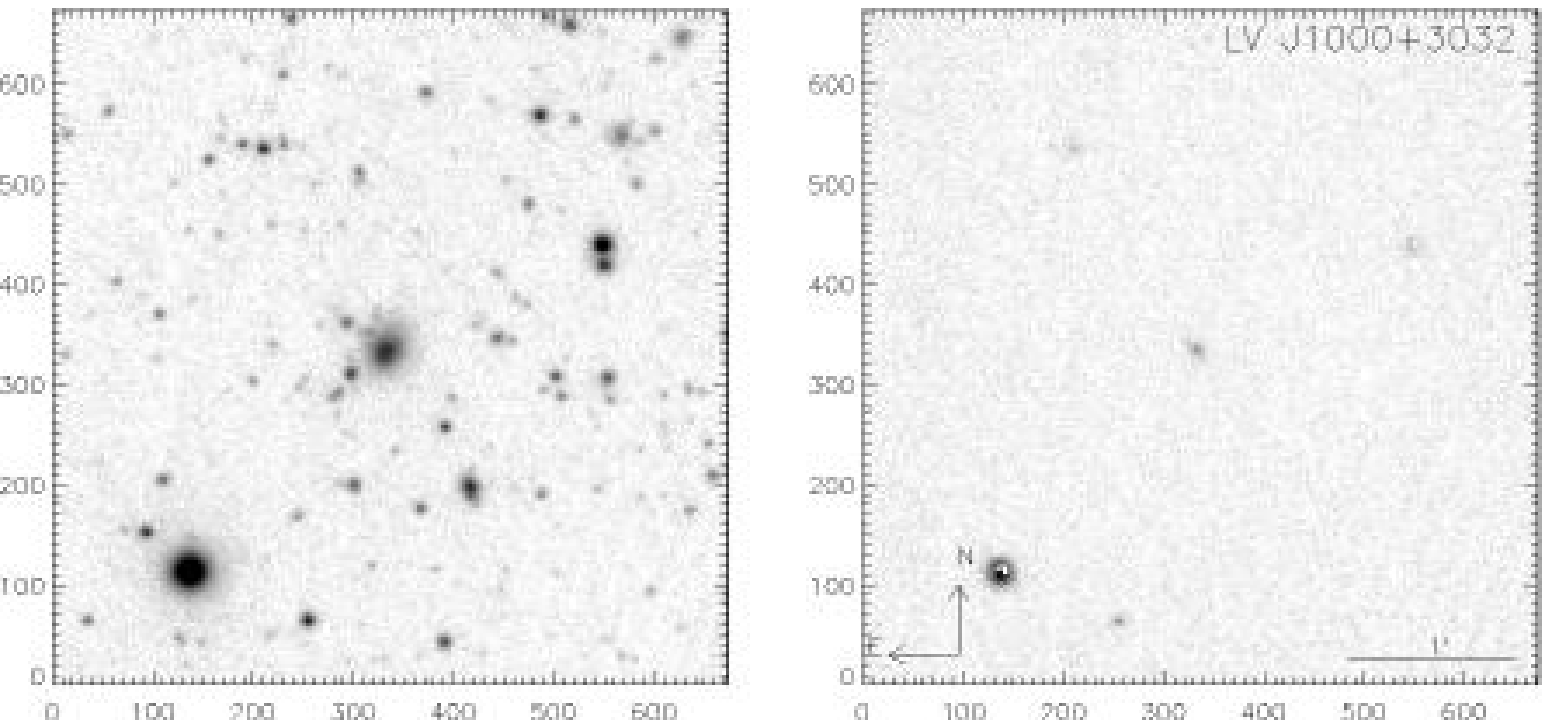}}

{\vbox {\vspace{6mm}
\centerline{\includegraphics[angle=0,width=0.8\textwidth,clip]{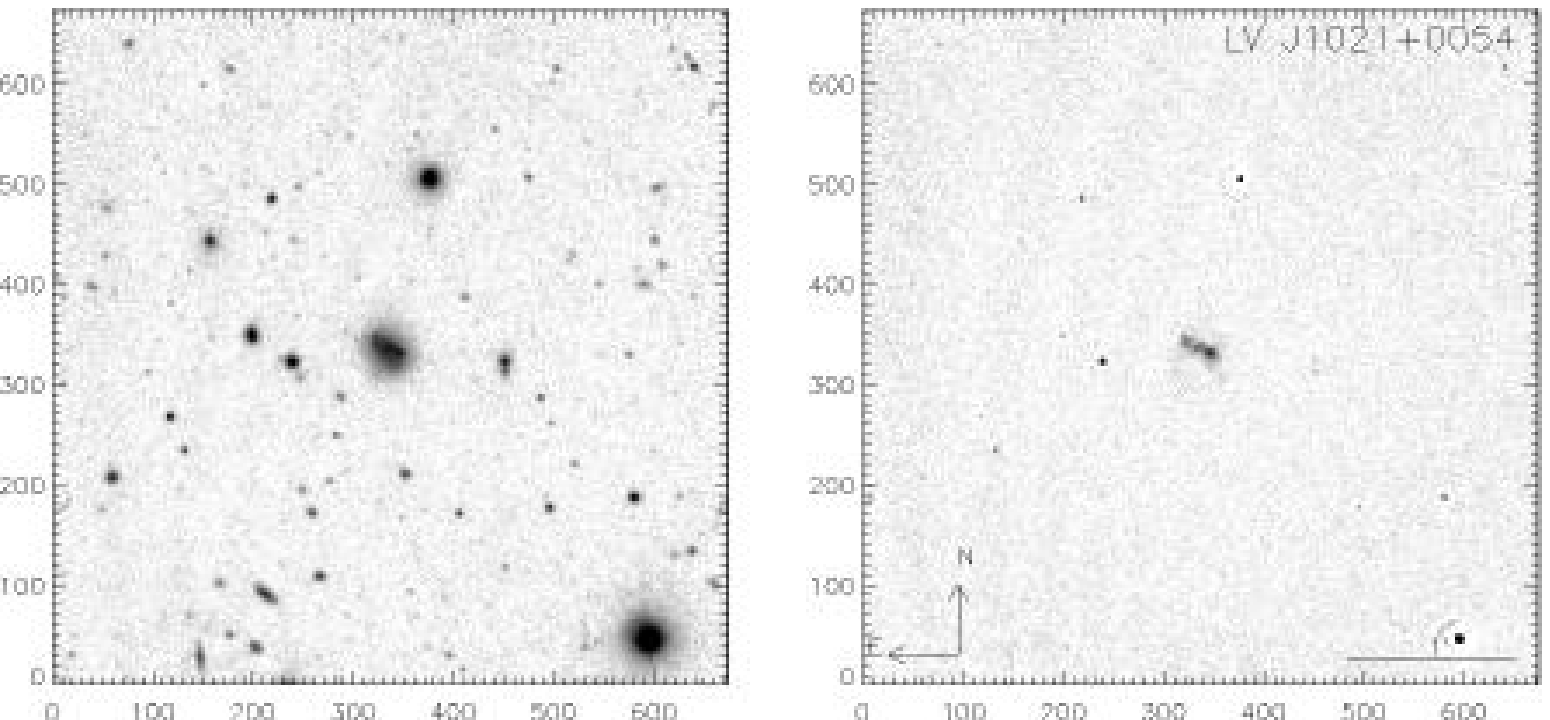}}}}
\centerline{\includegraphics[angle=0,width=0.8\textwidth,clip]{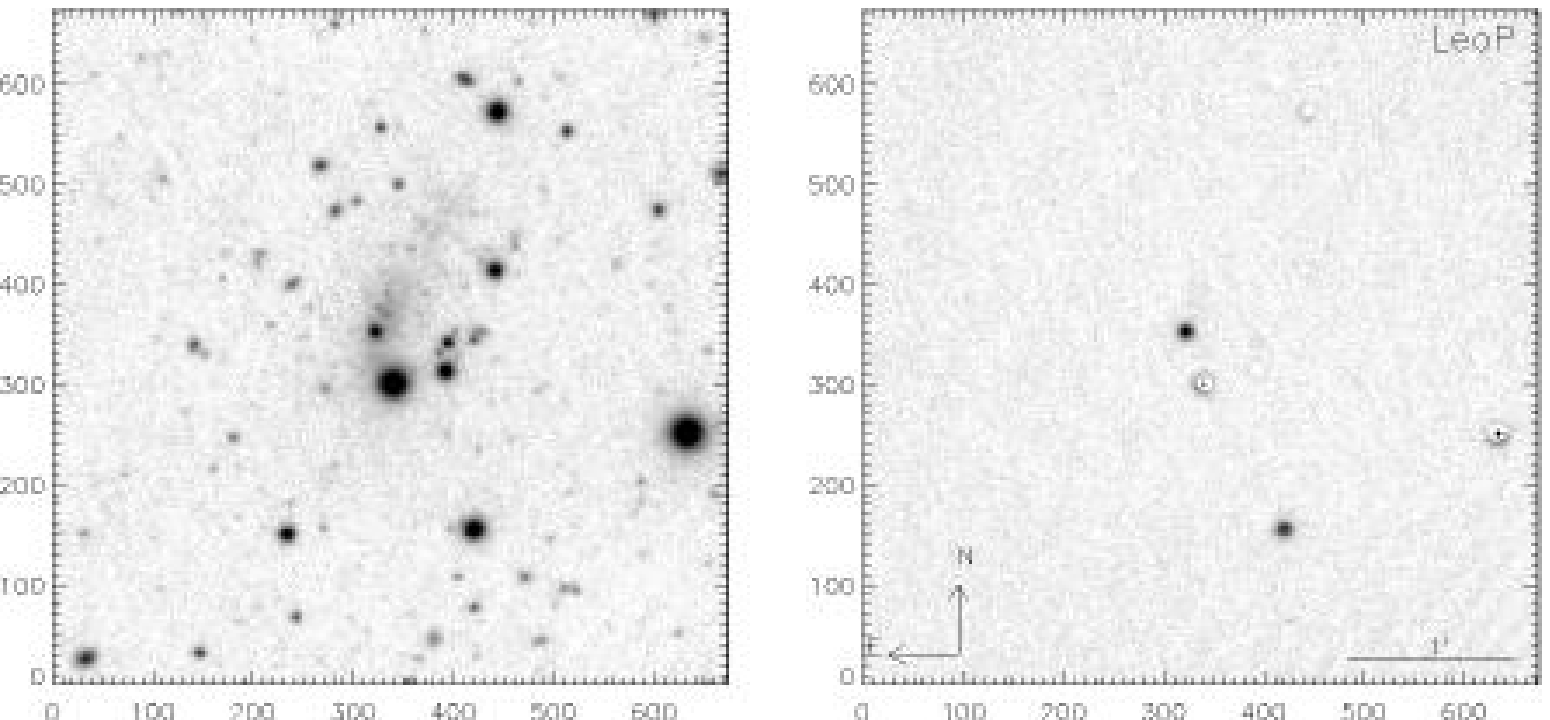}}
\centerline{\includegraphics[angle=0,width=0.8\textwidth,clip]{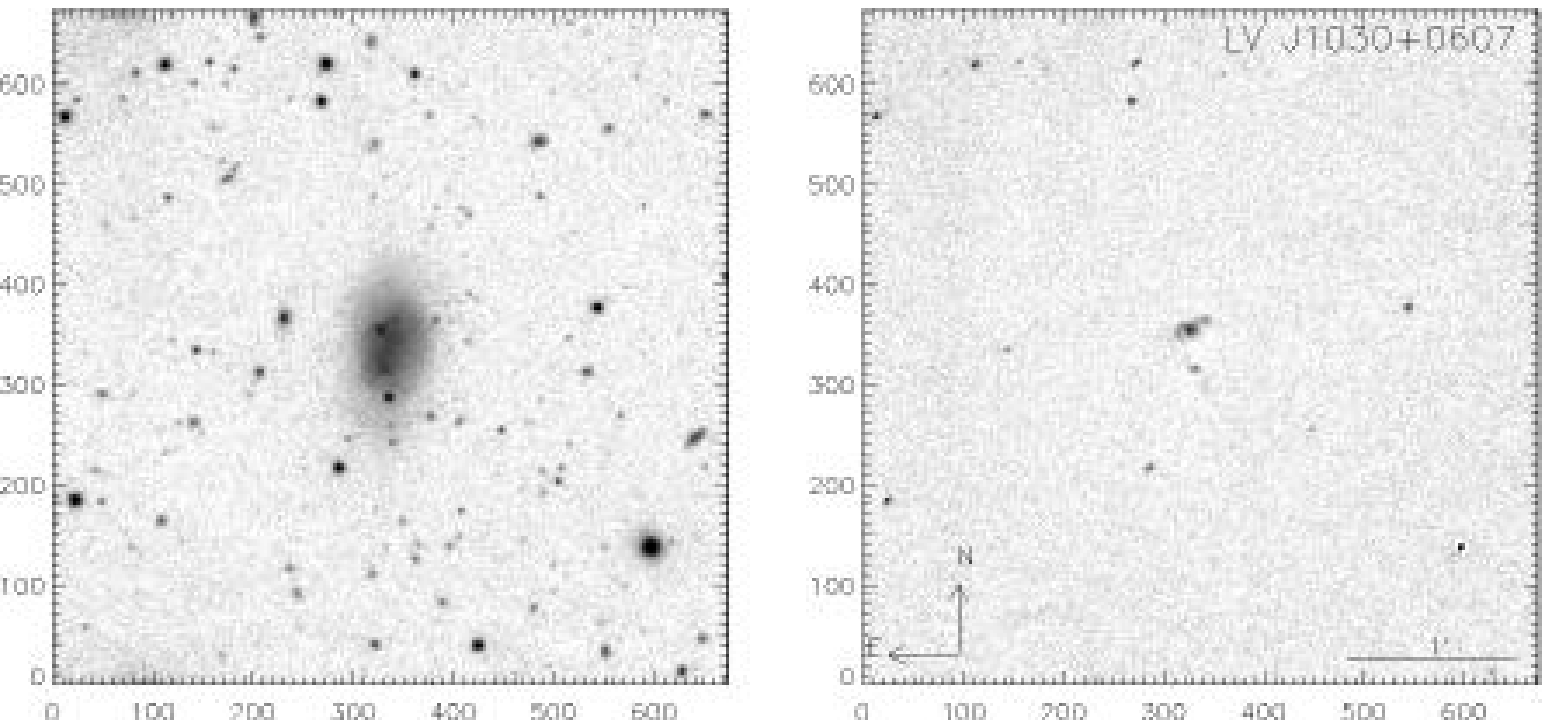}}

{\vbox {\vspace{6mm}
\centerline{\includegraphics[angle=0,width=0.8\textwidth,clip]{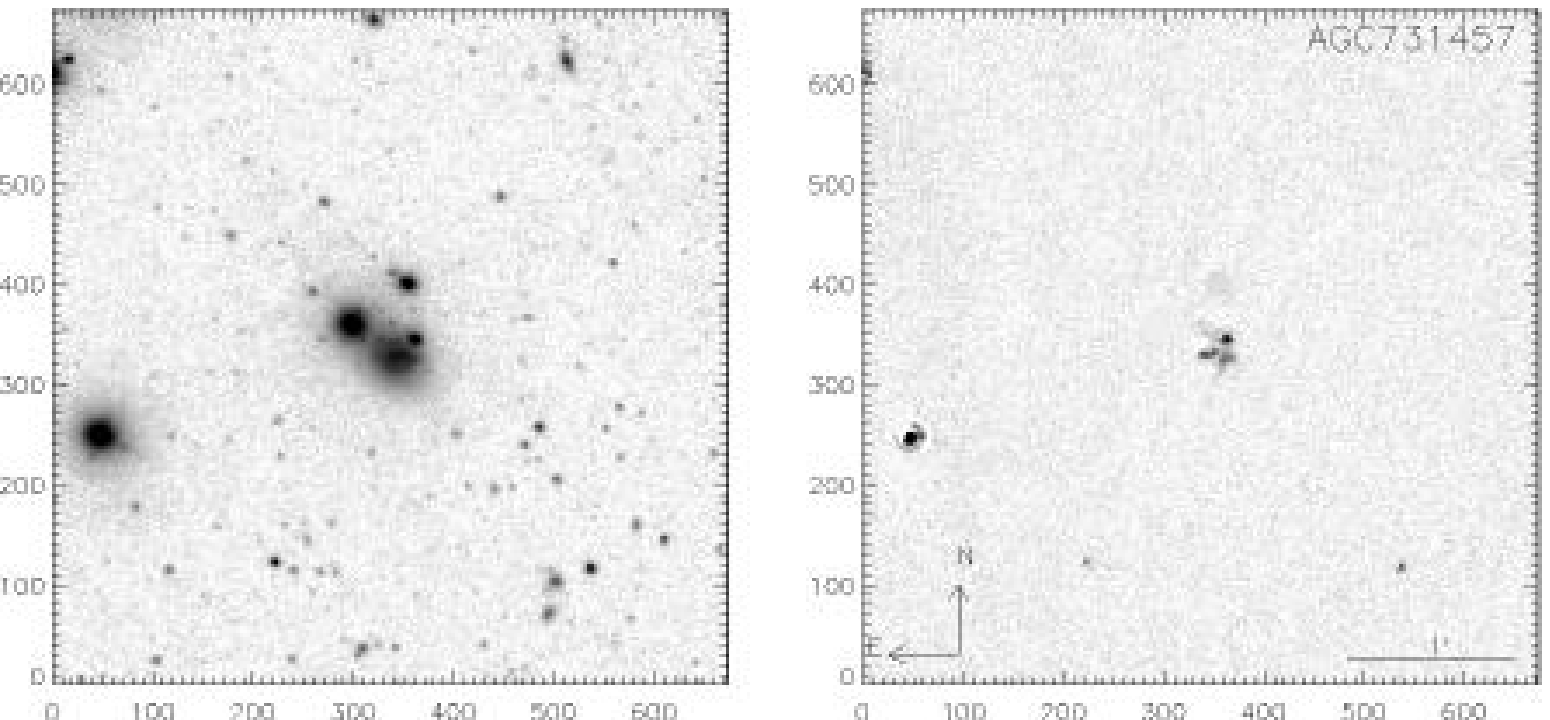}}}}
\centerline{\includegraphics[angle=0,width=0.8\textwidth,clip]{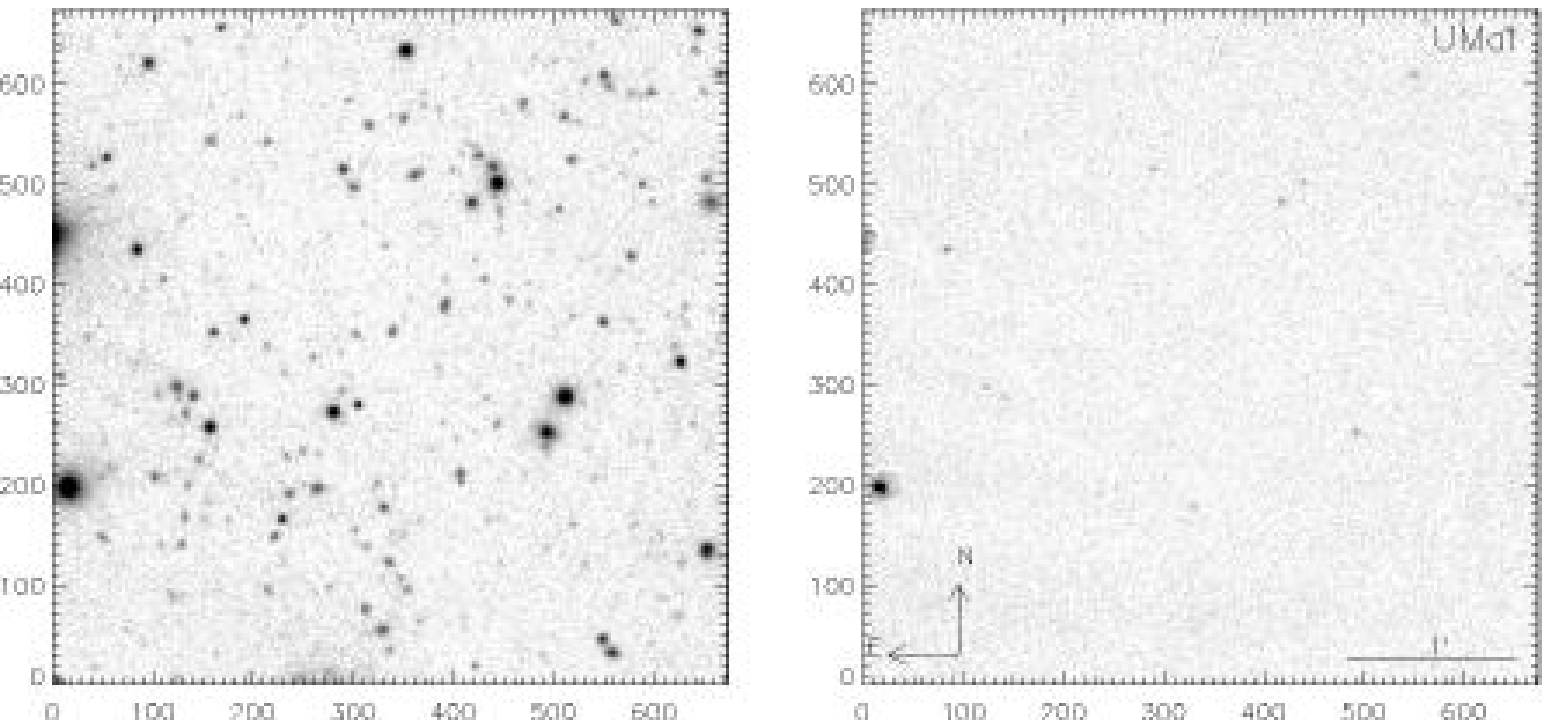}}
\centerline{\includegraphics[angle=0,width=0.8\textwidth,clip]{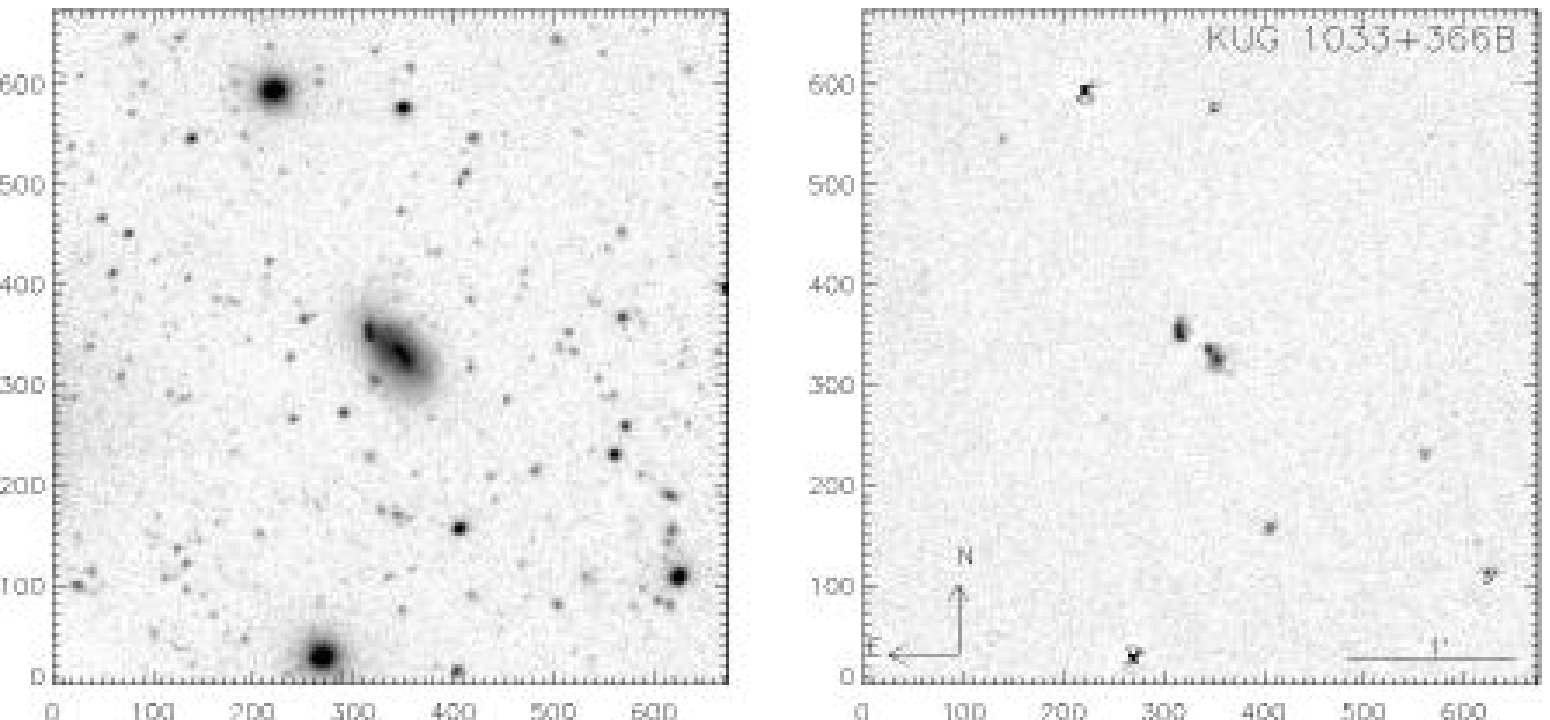}}

{\vbox {\vspace{6mm}
\centerline{\includegraphics[angle=0,width=0.8\textwidth,clip]{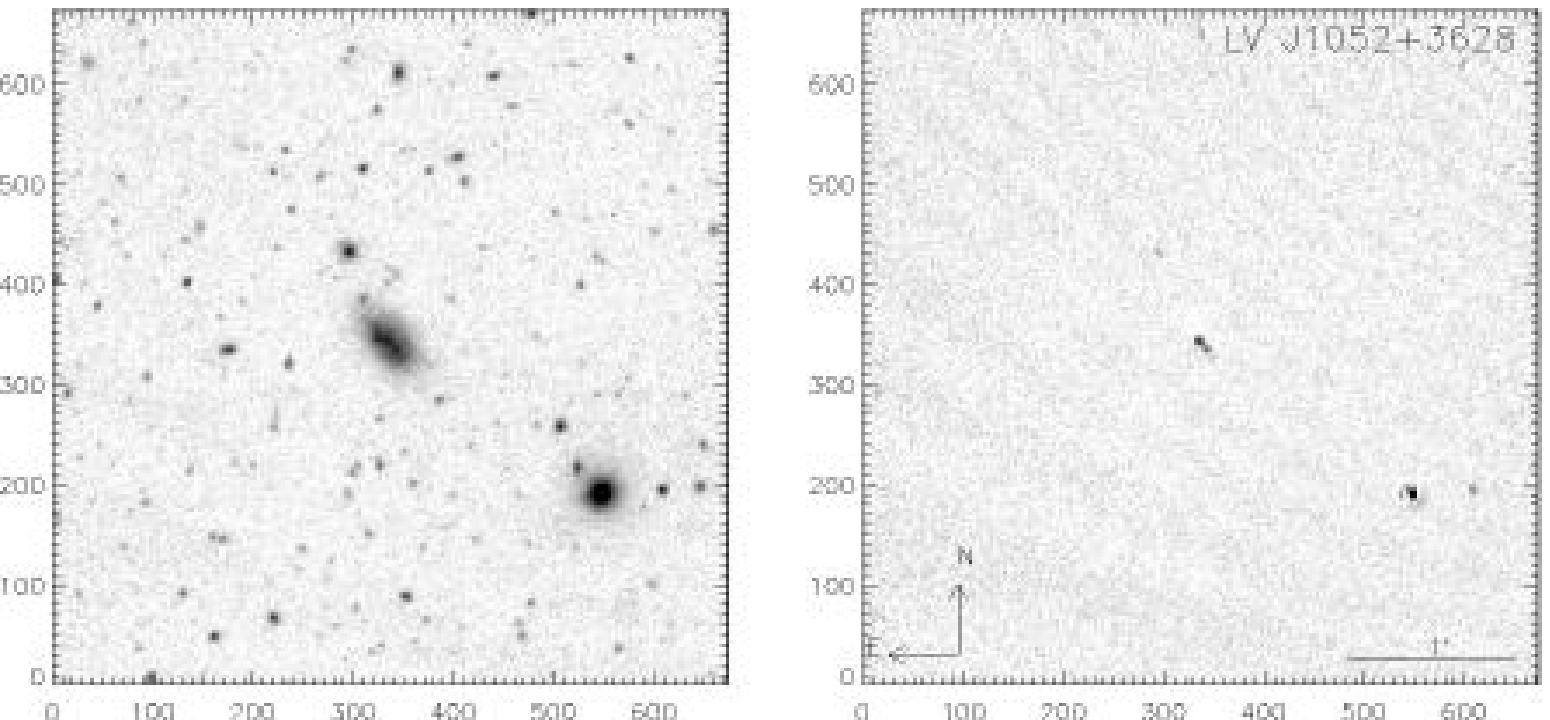}}}}
\centerline{\includegraphics[angle=0,width=0.8\textwidth,clip]{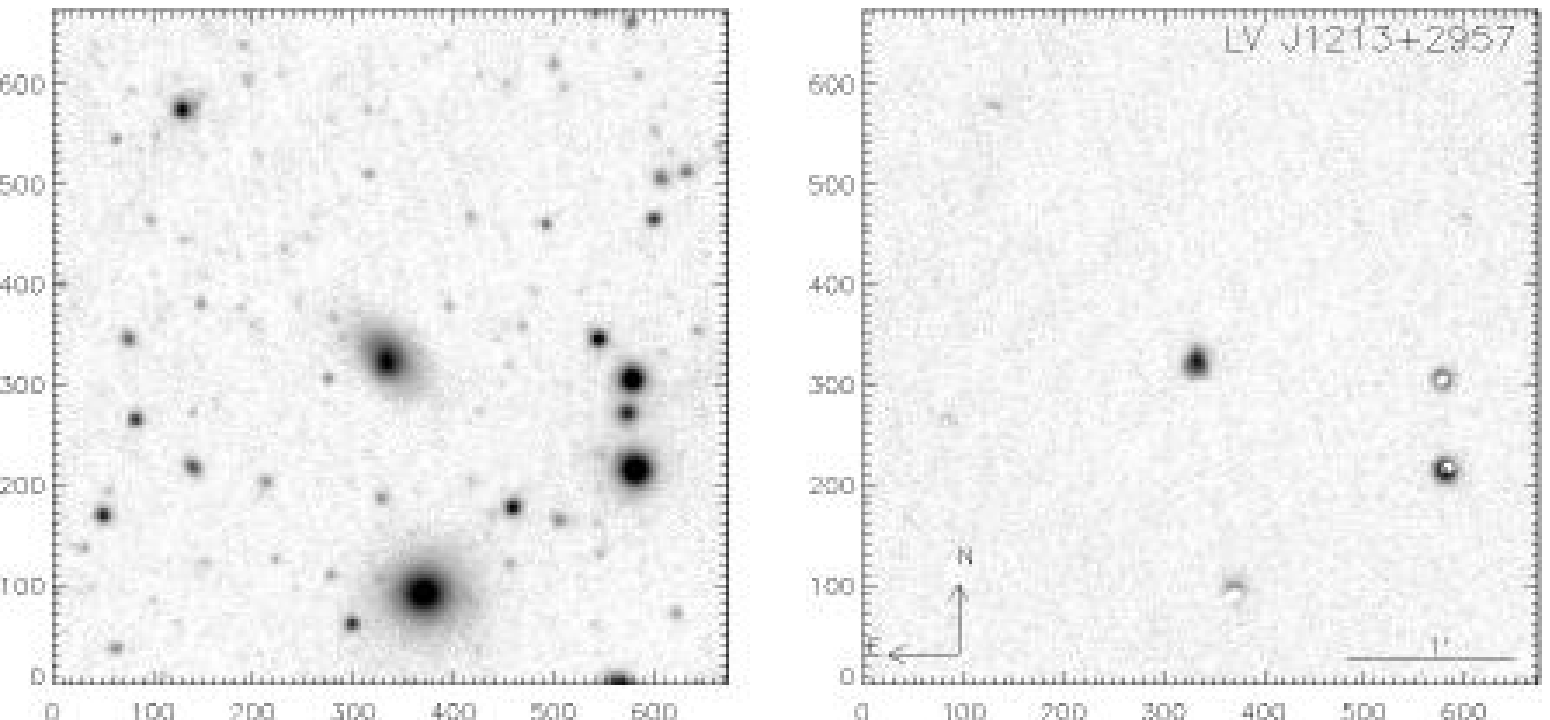}}
\centerline{\includegraphics[angle=0,width=0.8\textwidth,clip]{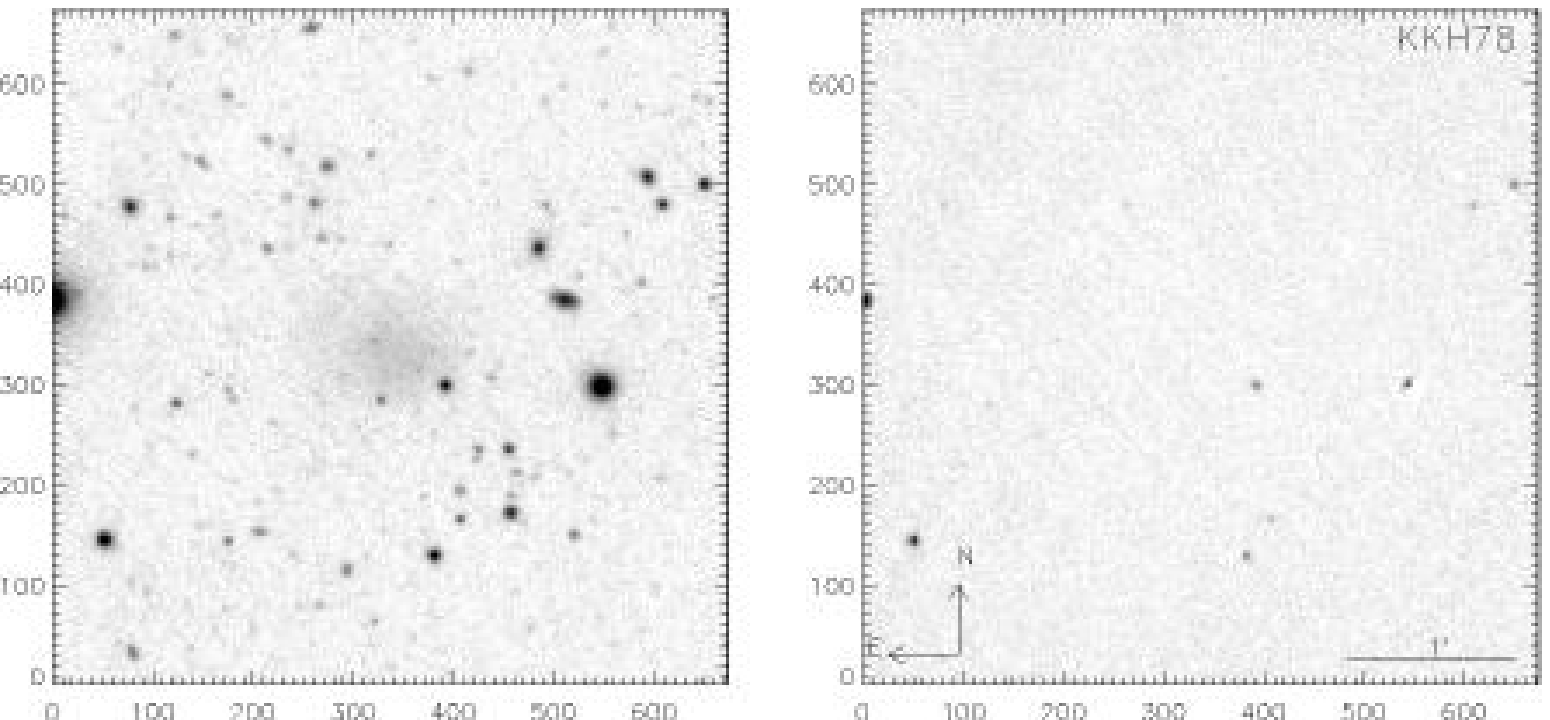}}

{\vbox {\vspace{6mm}
\centerline{\includegraphics[angle=0,width=0.8\textwidth,clip]{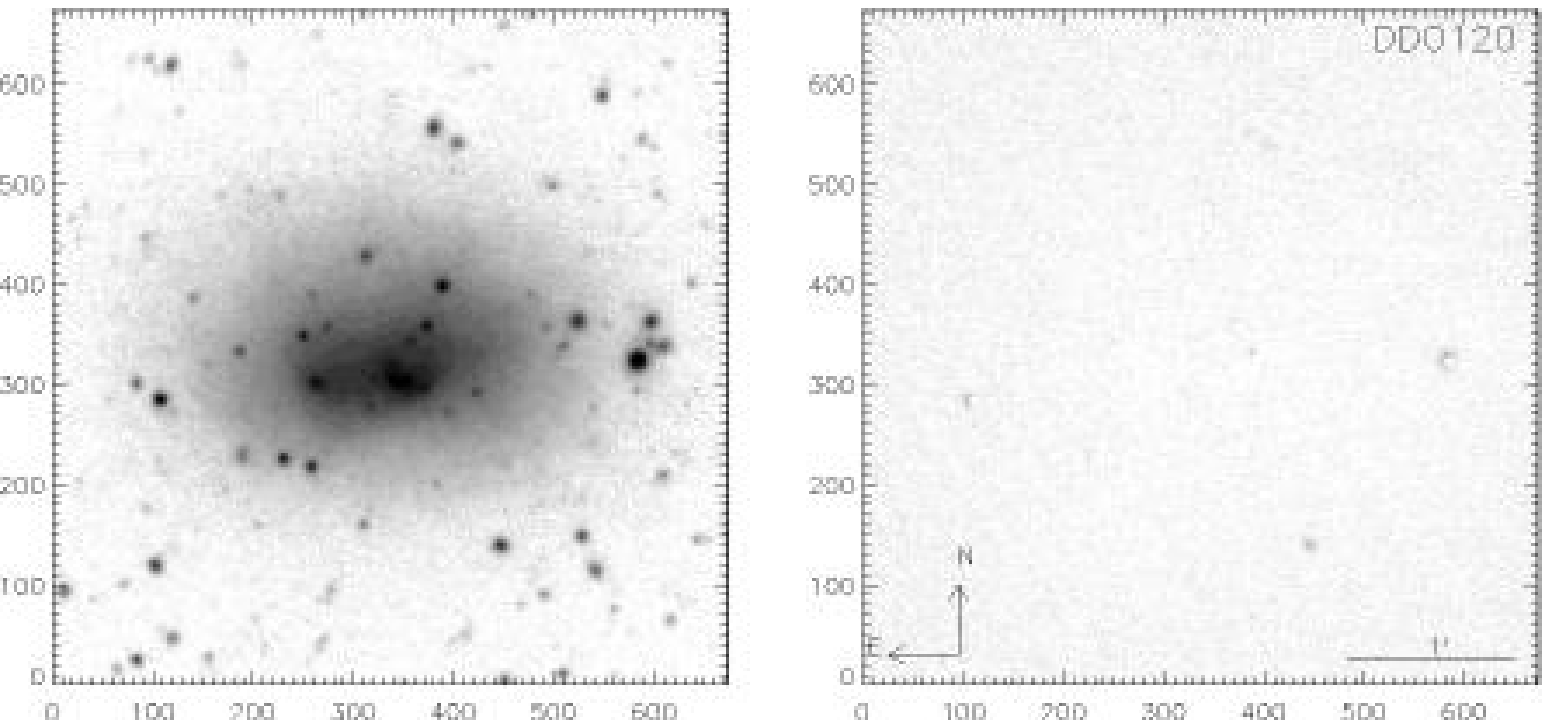}}}}
\centerline{\includegraphics[angle=0,width=0.8\textwidth,clip]{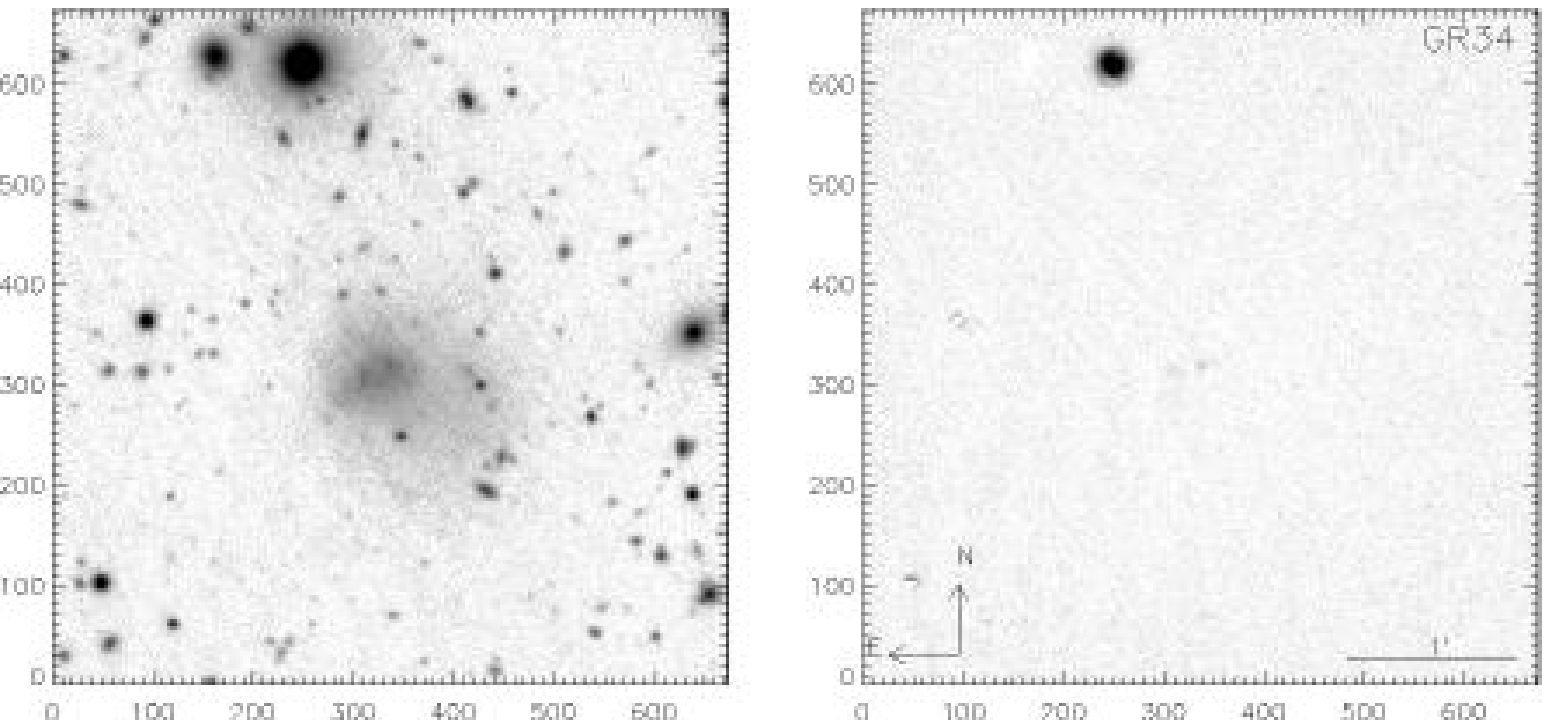}}
\centerline{\includegraphics[angle=0,width=0.8\textwidth,clip]{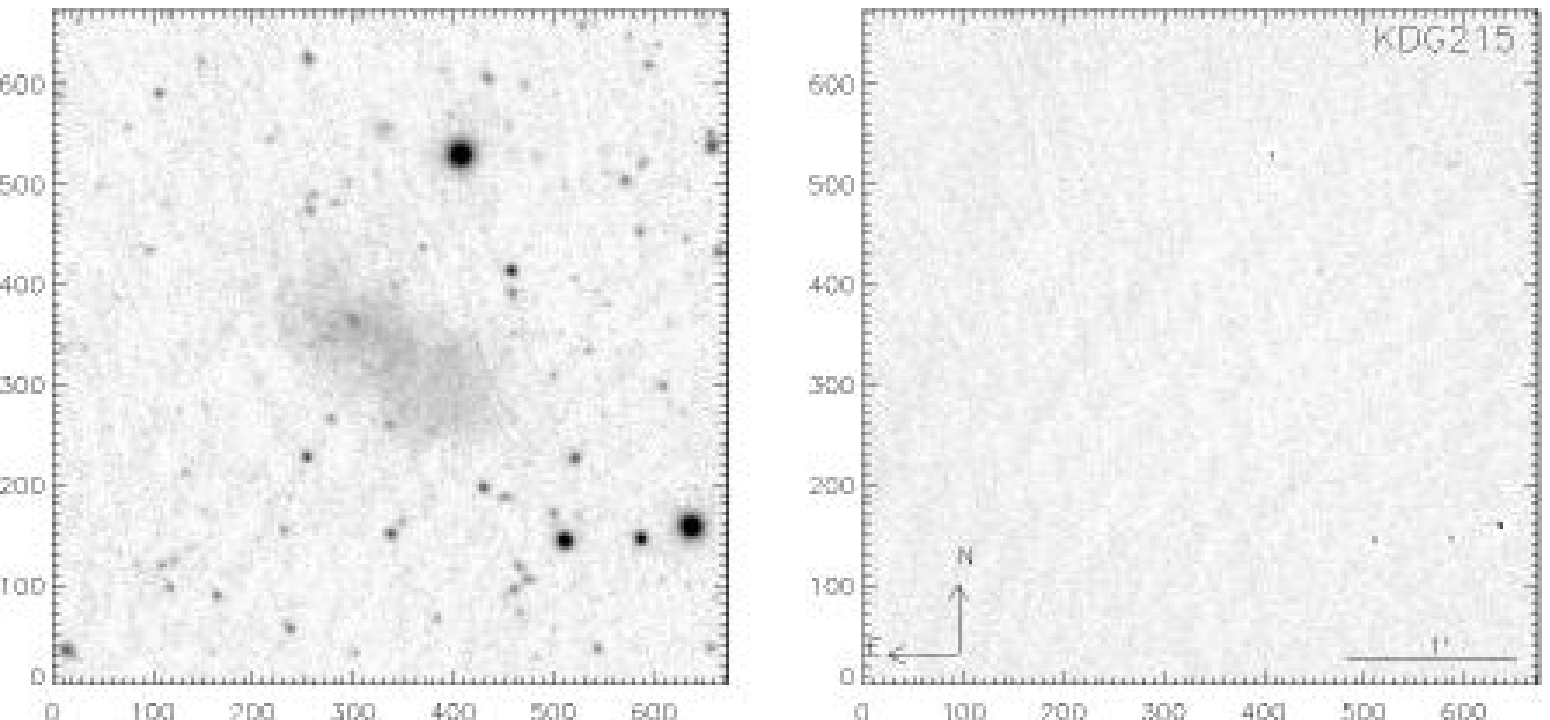}}

{\vbox {\vspace{6mm}
\centerline{\includegraphics[angle=0,width=0.8\textwidth,clip]{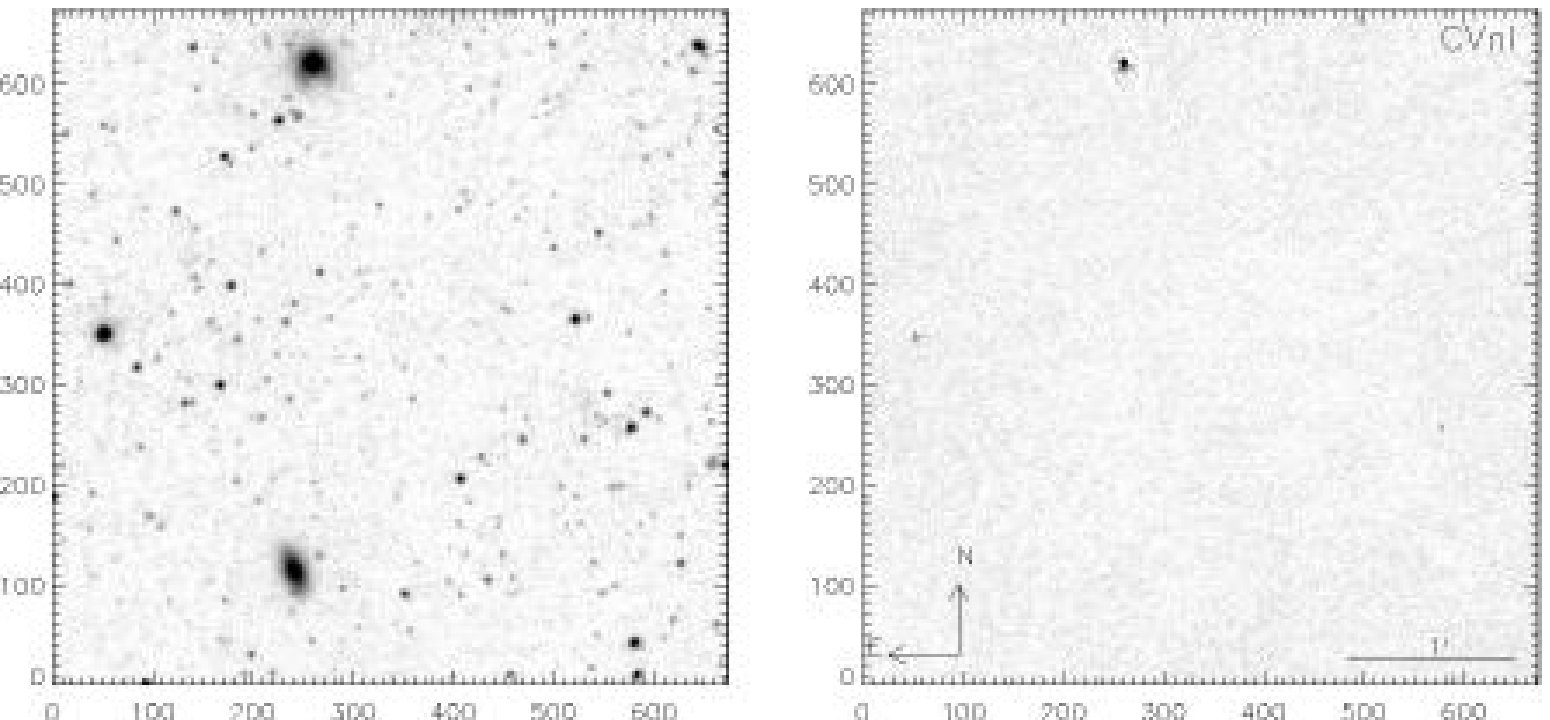}}}}
\centerline{\includegraphics[angle=0,width=0.8\textwidth,clip]{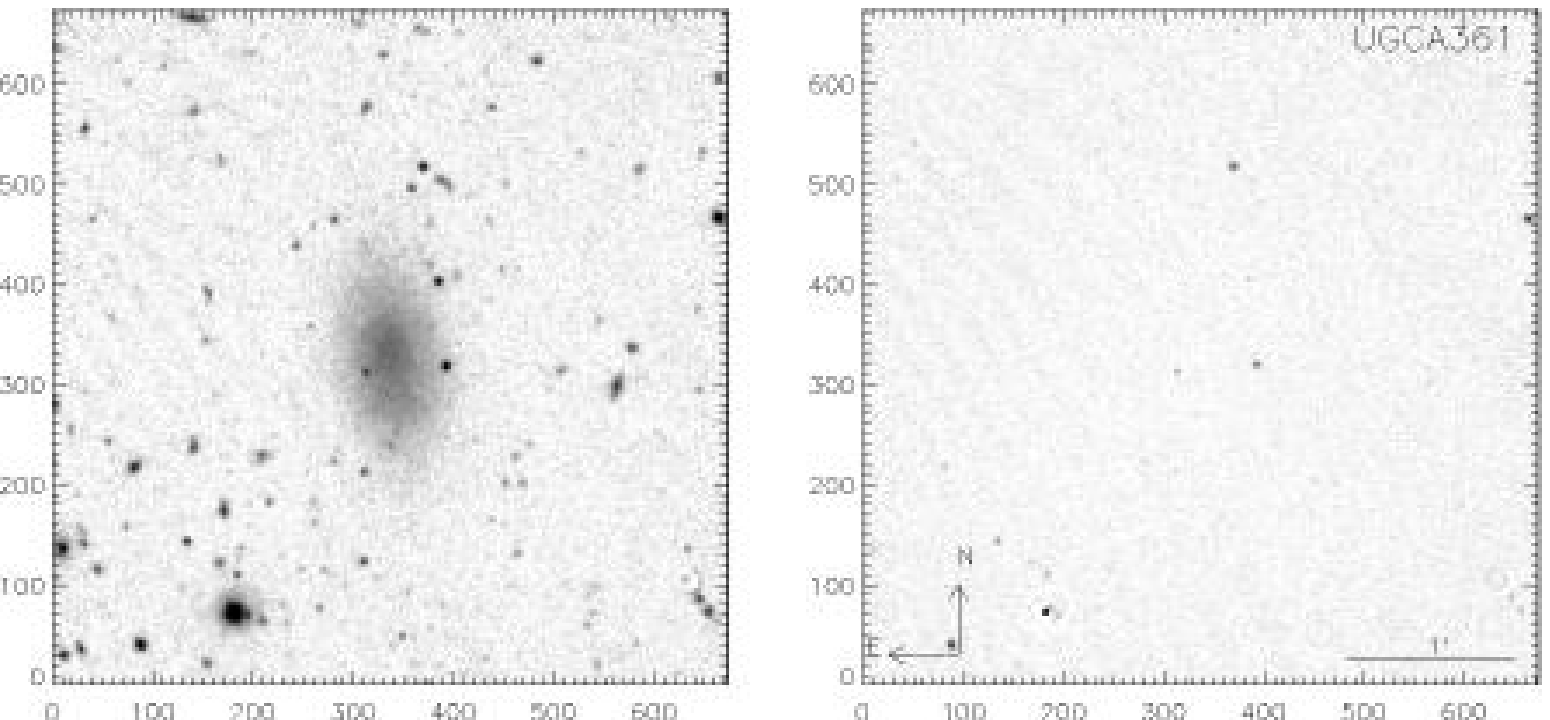}}
\centerline{\includegraphics[angle=0,width=0.8\textwidth,clip]{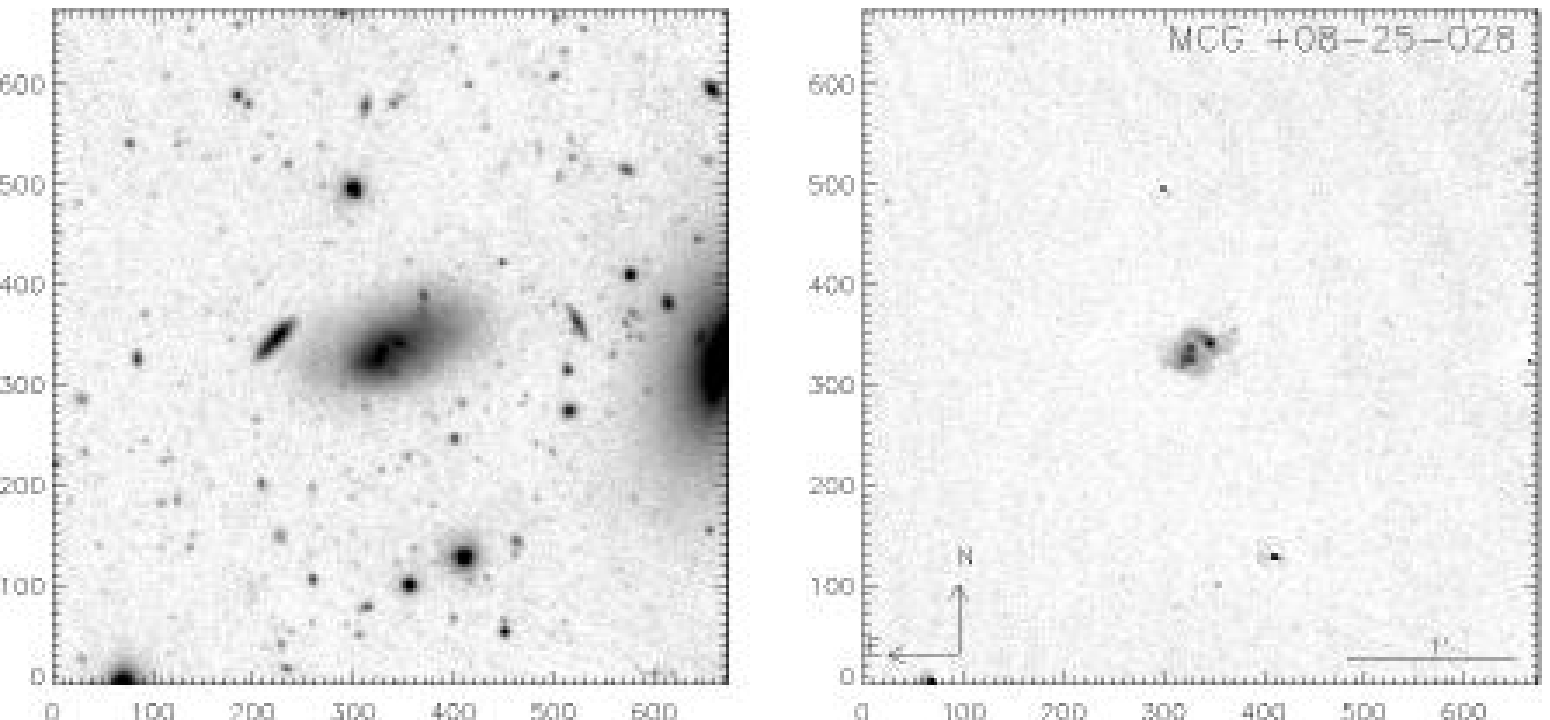}}

{\vbox {\vspace{6mm}
\centerline{\includegraphics[angle=0,width=0.8\textwidth,clip]{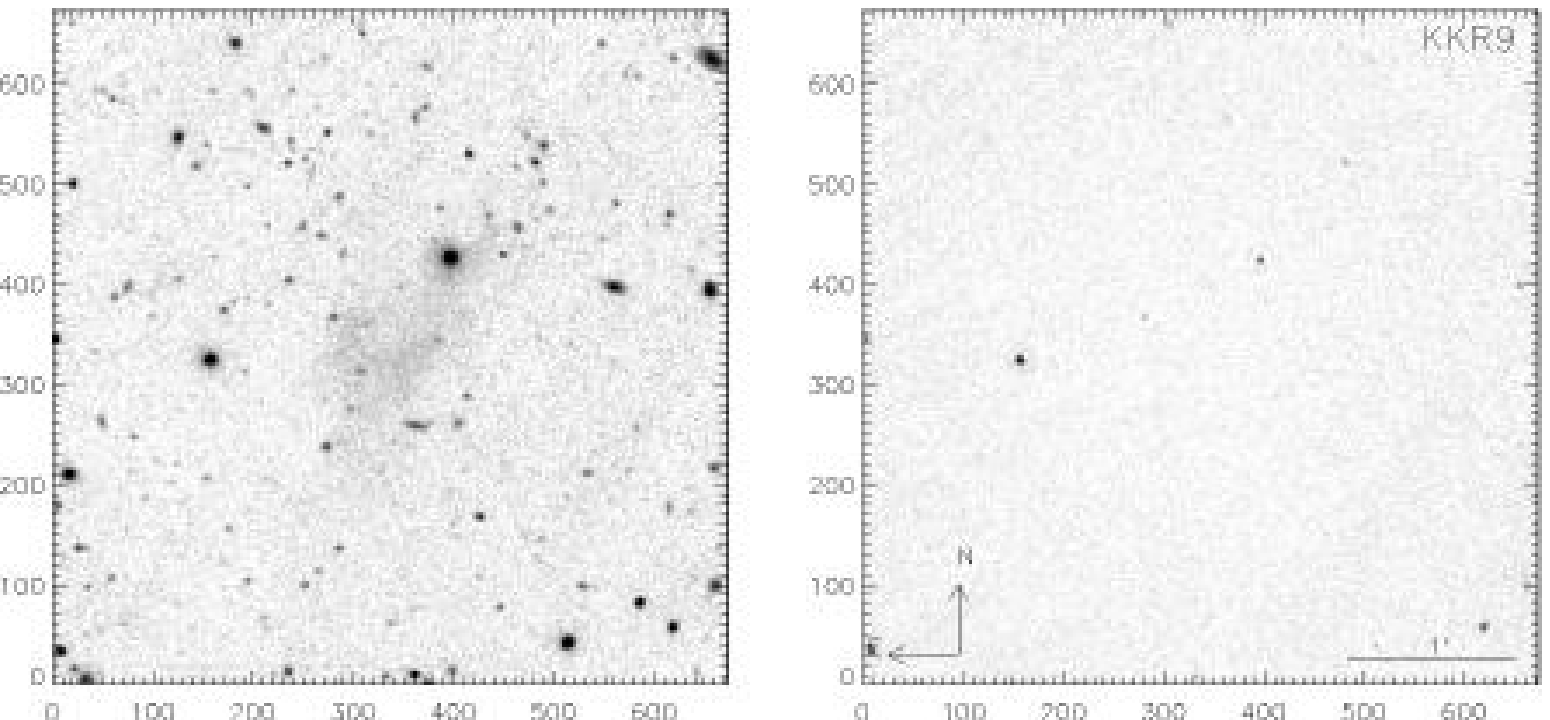}}}}
\centerline{\includegraphics[angle=0,width=0.8\textwidth,clip]{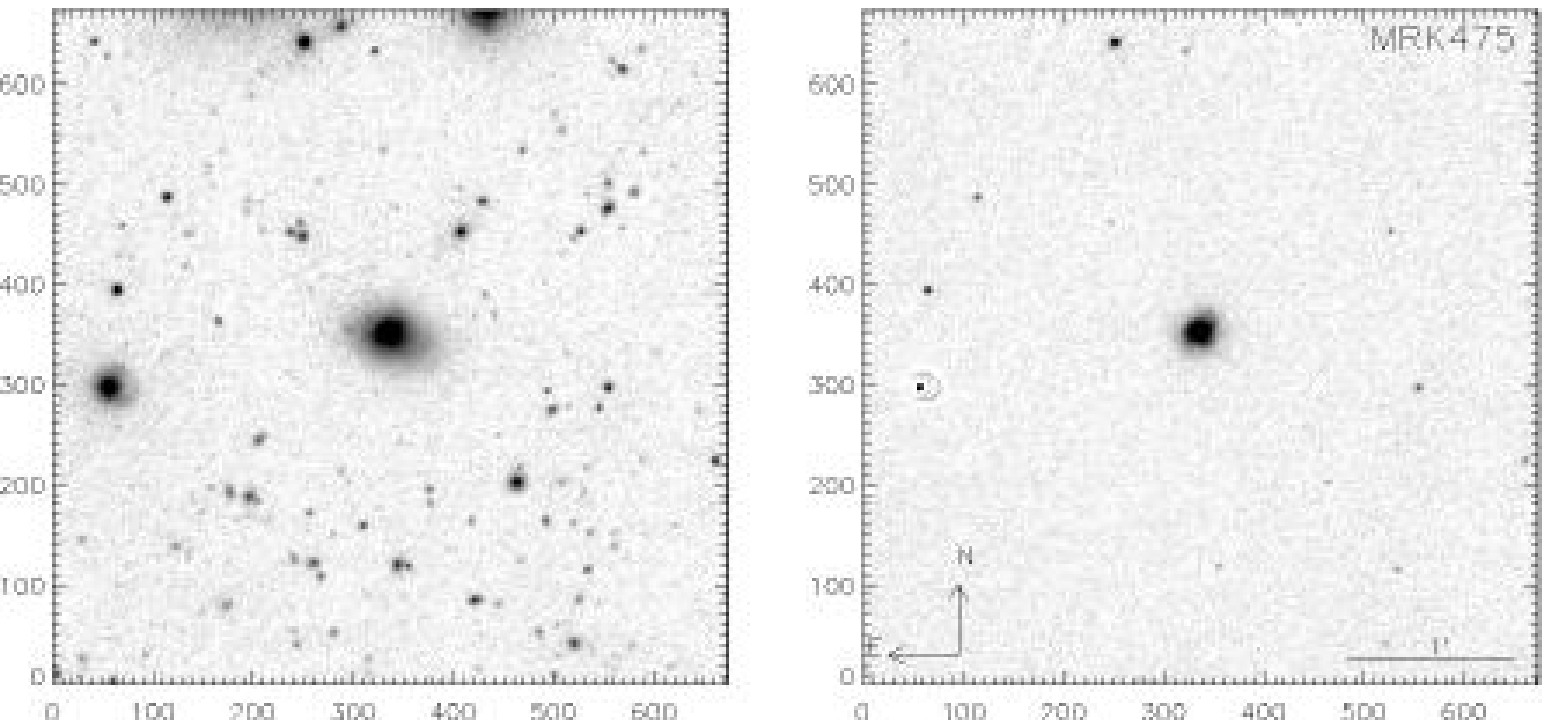}}
\centerline{\includegraphics[angle=0,width=0.8\textwidth,clip]{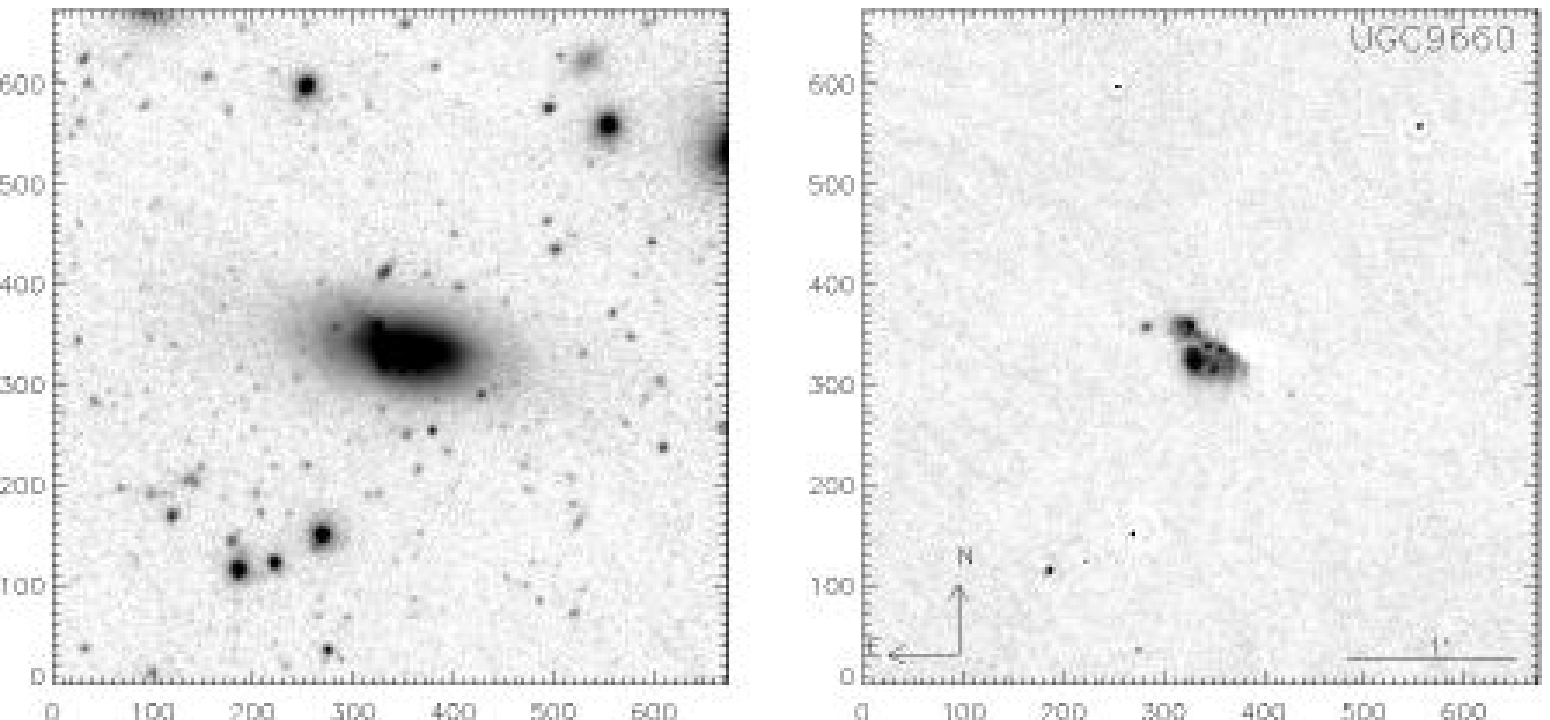}}

{\vbox {\vspace{6mm}
\centerline{\includegraphics[angle=0,width=0.8\textwidth,clip]{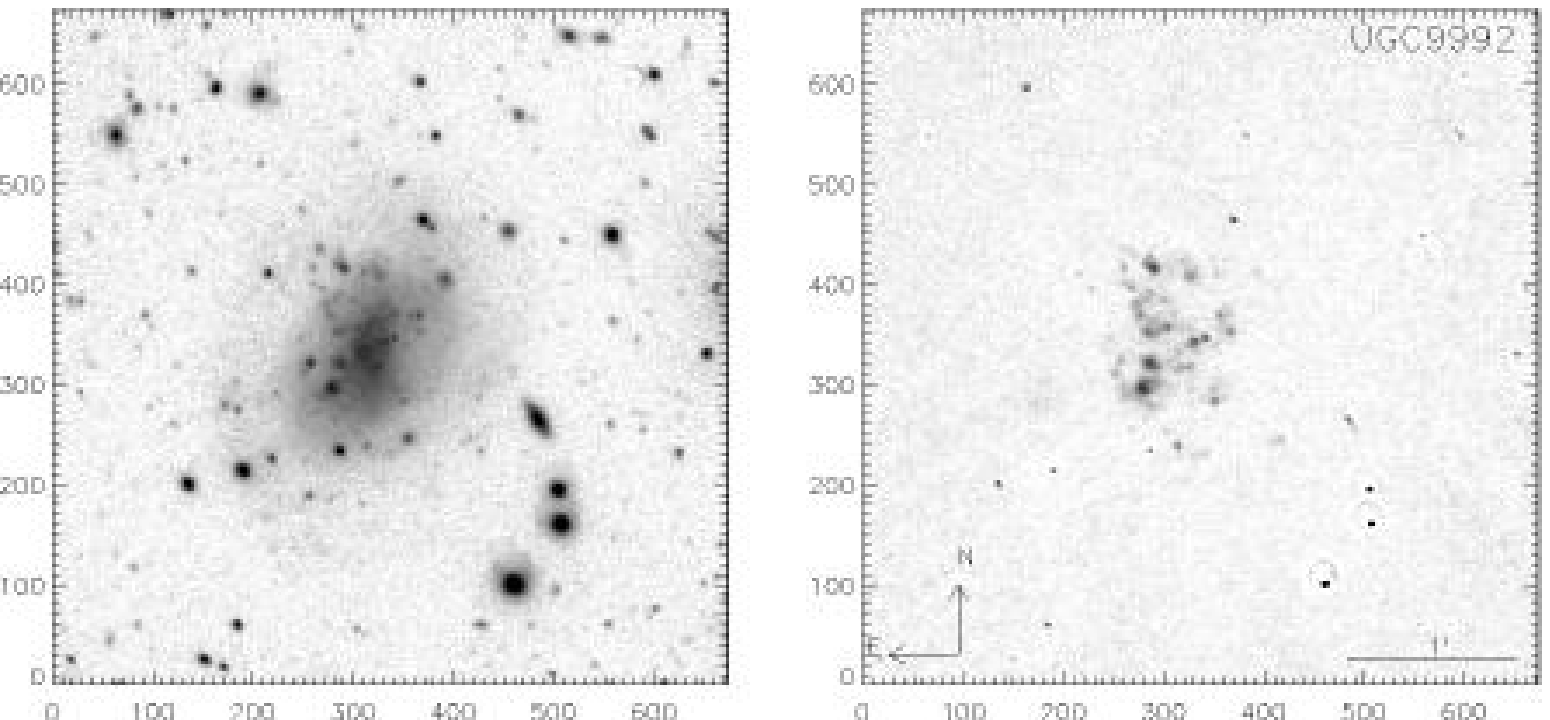}\vspace{10mm}}}}
\centerline{\includegraphics[angle=0,width=0.8\textwidth,clip]{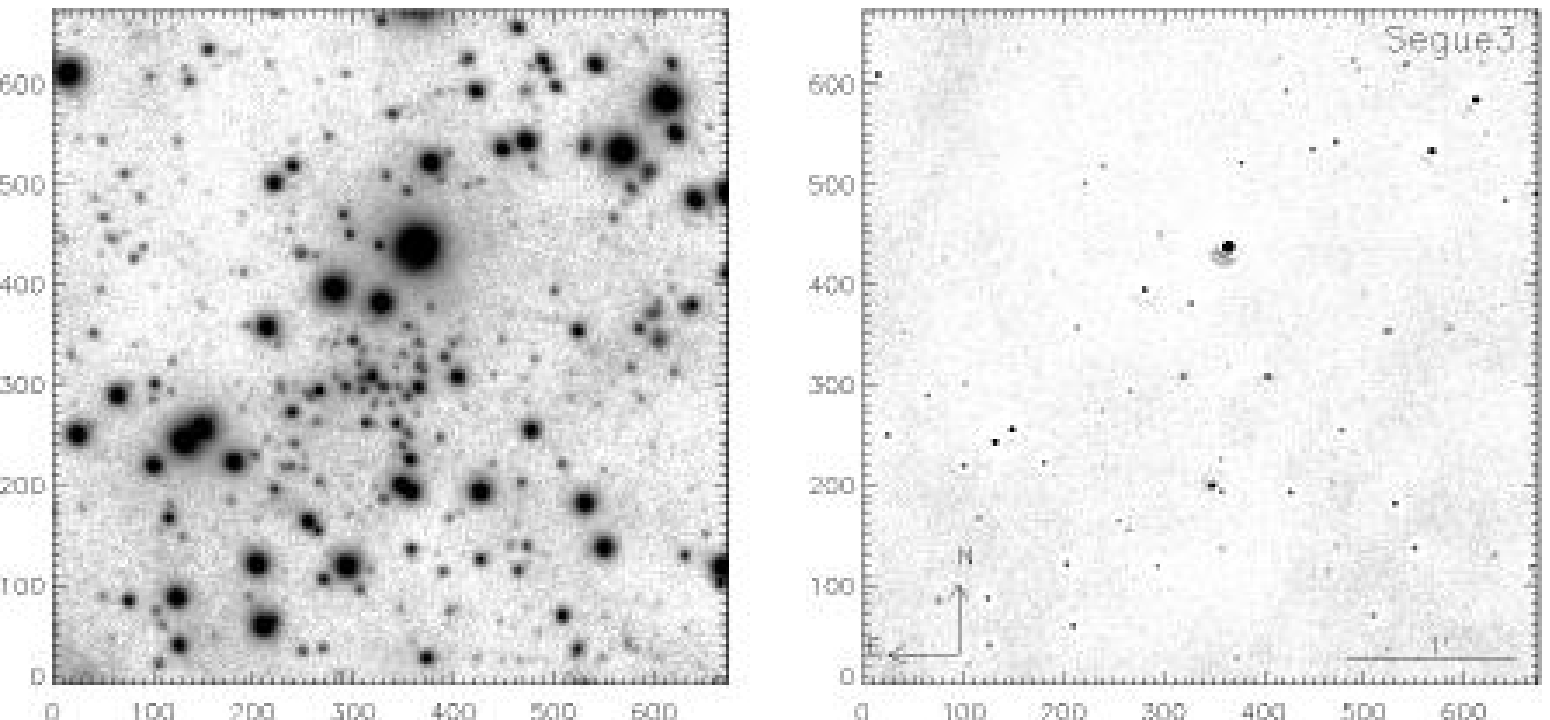}\vspace{10mm}}
\centerline{\includegraphics[angle=0,width=0.8\textwidth,clip]{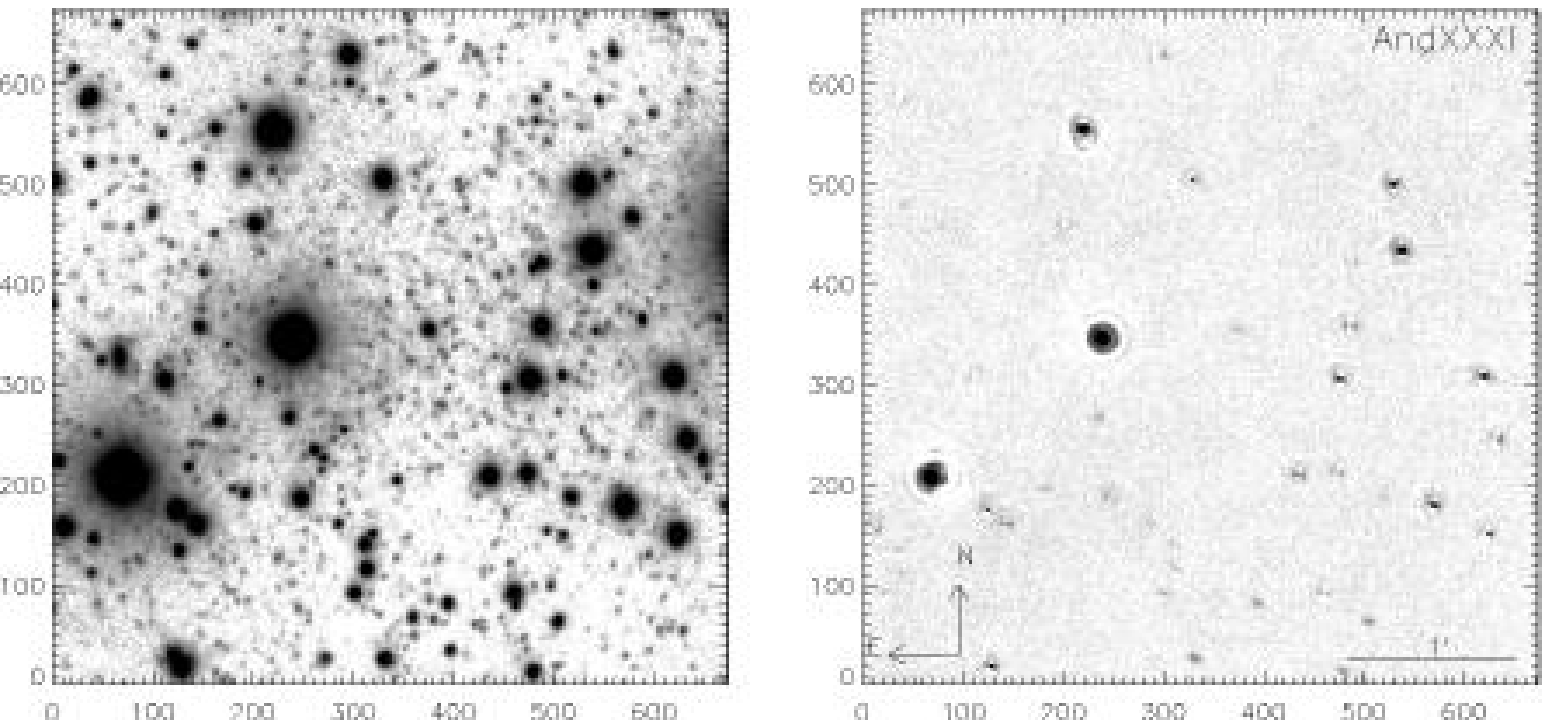}\vspace{40mm}}

\twocolumngrid


\begin{thebibliography}{}


\bibitem{1}I.~D.~Karachentsev, S.~S.~Kaisin, Z.~Tsvetanov, and H.~Ford,
\aaa {\bf 434}, 935 (2005).

\bibitem{2}S.~S.~Kaisin and I.~D. Karachentsev, Astrophysics {\bf 49},287 (2006).

\bibitem{3}I.~D.~Karachentsev and S.~S.~Kaisin, \aj {\bf 133}, 1883 (2007).

\bibitem{4}S.~S.~Kaisin and I.~D.~Karachentsev, \aaa {\bf 479}, 603 (2008).

\bibitem{5}I.~D.~Karachentsev and S.~S.~Kaisin, \aj {\bf  140}, 1241 (2010).

\bibitem{6}S.~S.~Kaisin, I.~D.~Karachentsev, and E.~I.~Kaisina, Astrophysics {\bf 54}, 315 (2011).

\bibitem{7}S.~S.~Kaisin and I.~D.~Karachentsev, Astrophysics {\bf 56}, 305 (2013).

\bibitem{8}S.~S.~Kaisin and I.~D.~Karachentsev,  \ab {\bf 68}, 381 (2013).

\bibitem{9}I.~D.~Karachentsev, D.~I.~Makarov, and E.~I.~Kaisina, \aj {\bf 145}, 101 (2013).

\bibitem{10}E.~I.~Kaisina, D.~I.~Makarov, I.~D.~Karachentsev, and S.~S.~Kaisin, \ab {\bf 67}, 115, (2012).

\bibitem{11}V.~L.~Afanasiev and A.~V.~Moiseev, Astronomy Lettets {\bf 31}, 194 (2005).

\bibitem{12}J.~B.~Oke, \aj {\bf 99}, 1621 (1990).

\bibitem{13}R.~C.~Kennicutt, \araa {\bf 36}, 189 (1998).

\bibitem{14}D.~J.~Schlegel, D.~P.~Finkbeiner, and M.~Davis, \apj {\bf 500}, 525 (1998).

\bibitem{15}M.~A.~W.~Verheijen,  \apj {\bf 563}, 694 (2001).

\bibitem{16}J.~C.~Lee, R.~C.~Kennicutt, J.~G.~Funes, et al., \apj {\bf 692}, 1305 (2009).

\bibitem{17}A.~Gil~de~Paz, S.~Boissier, B.~F.~ Madore,  et al., 2007, \apjs {\bf 173}, 185 (2007).

\bibitem{18}N.~F.~Martin, C.~T.~Slater, E.~F.~Schlafly, et al., \apj {\bf 772}, 15 (2013).

\bibitem{19}N.~F.~Martin, E.~F.~Schlafly,  C.~T.~Slater, et al.,  \apj {\bf 779L}, 10 (2013).

\bibitem{20}A.~D.~Mackey, A.~P.~Huxor, N.~F.~Martin, et al.,  \apj {\bf 770L}, 17 (2013).

\bibitem{21}D.~Martinez-Delgado, J.~Fliri, R.~Laesker, et al. (in preparation).

\bibitem{22}V.~Belokurov, M.~G.~Walker, N.~W.~Evans, et al.,  \mnras {\bf 397}, 1748 (2009).

\bibitem{23}B.~Willman, J.~J.~Dalcanton,  D.~Martinez-Delgado, et al.,  \apj {\bf 626L}, 85 (2005).

\bibitem{24}D.~B.~Zucker, V.~Belokurov, N.~W.~Evans,  et al., \apj  {\bf 643L}, 103 (2006).

\bibitem{25}V.~Belokurov, M.~G.~Walker, N.~W.~Evans,  et al.,  \apj {\bf 712L}, 103 (2010).

\bibitem{26}J.~B.~Jensen, J.~L.~Tonry, B.~J.~Barris,  et al.,  \apj {\bf 583}, 712 (2003).

\bibitem{27}B.~S.~Koribalski, L.~Staveley-Smith, V.~Kilborn, et al.,  \aj {\bf 128}, 16 (2004).

\bibitem{28}E.~Bernstein-Cooper, J.~M.~Cannon, and E.~A.~Adams, \aj {\bf 148}, 35 (2014).

\bibitem{29}K.~N.~Abazajian, J.~K.~Adelman-McCarthy,  M.~A.~Ag\"{u}eros, et al.  \apjs {\bf 182}, 543 (2009).

\bibitem{30}I.~D.~Karachentsev, O.~G.~Nasonova, and H.~M.~Courtois,  \apj {\bf 743}, 123 (2011).

\bibitem{31}I.~D.~Karachentsev, R.~B.~Tully, P.~F.~Wu, et al.,  \aj {\bf 782}, 4 (2014).

\bibitem{32}R.~B.~Tully and J.~R.~Fisher, 1977, \aaa {\bf 54}, 661 (1977).

\bibitem{33}J.~Pflamm-Altenburg, C.~Weidner, and P.~Kroupa, \apj {\bf 671}, 1550 (2007).

\bibitem{34}I.~D.~Karachentsev and E.~I.~Kaisina, \aj {\bf 146}, 46 (2013).

\bibitem{35}I.~D.~Karachentsev, V.~E.~Karachentseva, O.~V.~Melnyk, and H.~M.~Courtois, \ab {\bf 68}, 243 (2013).

\end{thebibliography}
\end{document}